\documentclass[reprint,prb,footinbib]{revtex4-2}
\usepackage{graphicx}
\usepackage{dcolumn}
\usepackage{bm}
\usepackage{amsmath,amsfonts,amssymb,braket,txfonts}
\renewcommand{\phi}{\varphi}
\usepackage[version=4]{mhchem}

\begin{document}

\title{Light-induced nonlinear Edelstein effect under ferroaxial ordering}
\author{Akimitsu Kirikoshi$^{1}$ and Satoru Hayami$^{2}$}
\affiliation{
  $^{1}$Research Institute for Interdisciplinary Science, Okayama University, Okayama 700-8530, Japan
  \\
  $^{2}$Graduate School of Science, Hokkaido University, Hokkaido 060-0810, Japan
}

\begin{abstract}
  Ferroaxial ordering, a spontaneous rotational distortion of the atomic arrangement, brings about a cross-product-type spin-orbit coupling (SOC) manifested as an electric toroidal (ET) dipole.
  We propose the light-induced nonlinear Edelstein effect (NLEE)---a second-order optical response in which a static magnetization is induced by a time-dependent electric field---as a promising probe of ferroaxial ordering.
  First, we elucidate the relationship between the NLEE tensor and the ET dipole.
  By decomposing the polarization modes of light, we find that both the linearly and circularly polarized light couple to the ET dipole via distinct mechanisms.
  We then demonstrate the NLEE using a minimal model that incorporates ferroaxial ordering.
  Our analysis reveals that effective coupling between orbital magnetization and SOC induces spin magnetization.
  The spin magnetization is tilted owing to the ET dipole; the tilt angle reflects the ratio between the ferroaxial-origin SOC and the relativistic SOC.
\end{abstract}

\maketitle

\section{Introduction}

Ordered states accompanied by spontaneous symmetry breaking enrich the functional properties of materials.
For example, unlike normal electrical conduction, the anomalous Hall effect
(AHE)~\cite{nagaosaAnomalousHallEffect2010,xiaoBerryPhaseEffects2010} enables dissipation-free current in magnetic systems.
While the AHE is typically associated with ferromagnetic order, it can also arise in certain antiferromagnetic orders characterized by higher-rank multipoles~\cite{chenAnomalousHallEffect2014,suzukiClusterMultipoleTheory2017}.
The inverse-triangular noncollinear order of \ce{Mn3Sn} can be described within the cluster-multipole framework as a magnetic octupole that transforms under the same irreducible representation as the magnetic dipole; this symmetry allows a large intrinsic AHE despite vanishingly small net magnetization~\cite{nakatsujiLargeAnomalousHall2015}.
The discovery of the giant AHE in the antiferromagnet \ce{Mn3Sn} has attracted considerable attention for next-generation spintronic applications.

In recent years, increasing attention has been directed toward multipole order(ing), which encompasses not only magnetic and orbital orders and their combined degrees of freedom, including site and bond in crystals~\cite{hayamiUnifiedDescriptionElectronic2024}.
Such ordering is of great interest because associated cross-correlated phenomena are expected to yield unecplored functionality.
A representative example is the magnetic toroidal (MT) dipole, which characterizes vortex-like magnetic structures and induces magnetization (dielectric polarization) in response to an electric (a magnetic) field.
This effect arises when both spatial inversion (SI) and time-reversal (TR) symmetries are broken and has been extensively studied, primarily in antiferromagnetic insulators~\cite{edererMicroscopicTheoryToroidal2007,spaldinToroidalMomentCondensedmatter2008,kopaevToroidalOrderingCrystals2009}.

Here, we focus on the response associated with another toroidal dipole, the electric toroidal (ET) dipole, which serves as a microscopic order parameter for ferroaxial ordering. 
Ferroaxial ordering occurs at a phase transition accompanied by a rotational distortion of the atomic arrangement and has been reported in materials such as \ce{Rb(FeMo4)2}~\cite{jinObservationFerrorotationalOrder2020,hayashidaPhaseTransitionDomain2021} and \ce{NiTiO3}~\cite{hayashidaPhaseTransitionDomain2021,hayashidaVisualizationFerroaxialDomains2020,yokotaThreedimensionalImagingFerroaxial2022,fangFerrorotationalSelectivityIlmenites2023}.
In contrast with other dipole orders like the MT dipole order and magnetic dipole (ferromagnetic) order, ferroaxial order with an ET dipole is difficult to control by static electromagnetic fields because both SI and TR symmetries are preserved.
From a theoretical perspective, the expression of the ET dipole has been formulated as~\cite{hayamiElectricFerroAxialMoment2022}
\begin{equation}
  \mathbf{G}=\mathbf{l}\times\mathbf{s},
  \label{eq:etd}
\end{equation}
where $\mathbf{l}$ and $\mathbf{s}$ are the orbital and spin angular momenta, respectively.
One of the authors recently showed on a cluster model that the ferroaxial phase transition gives rise to the cross-product-type spin-orbit coupling (SOC) described by Eq.~\eqref{eq:etd}~\cite{indaEmergentCrossproducttypeSpinorbit2025}.
A distinctive feature of ferroaxial systems is their off-diagonal response to conjugate fields;
proposed responses include spin-current generation~\cite{royUnconventionalSpinHall2022, hayamiElectricFerroAxialMoment2022}, antisymmetric thermopolarization~\cite{nasuAntisymmetricThermopolarizationElectric2022}, third-order transverse magnetization~\cite{indaNonlinearTransverseMagnetic2023}, and second-order nonlinear magnetostriction~\cite{kirikoshiRotationalResponseInduced2023}.
Very recentyly, domain-encoded nonlinear magnetic susceptibility consistent with the electric-toroidic/axial-multipole framework was reported in doped ilmenite \ce{FeTiO3}~\cite{duElectricToroidalInvariance2026}, providing experimental support for the transverse cross-correlations discussed here.
These point to new functionalities and provide indirect experimental avenues to identify ferroaxial ordering.

In the present paper, we investigate photomagnetization known as the light-induced nonlinear Edelstein effect (NLEE)~\cite{battiatoQuantumTheoryInverse2014,berrittaInitioTheoryCoherent2016,xuLightinducedStaticMagnetization2021,fregosoBulkPhotospinEffect2022} under ferroaxial ordering.
The light-induced NLEE denotes a static magnetization $\delta M$ generated by the second-order response to an oscillating electric field $E(\omega)$ with frequency $\omega$, satisfying $\delta M\propto E(\omega)E(-\omega)$.
Unlike the linear Edelstein effect (LEE)~\cite{edelsteinSpinPolarizationConduction1990}, which requires breaking of SI symmetry, the NLEE can occur even when both SI and TR symmetries are preserved.
Moreover, unlike the current-induced NLEE~\cite{xiaoIntrinsicNonlinearElectric2022,xiaoTimeReversalEvenNonlinearCurrent2023,guoExtrinsicContributionNonlinear2024,kodamaDirectObservationCurrentinduced2024,baekNonlinearOrbitalSpin2024,oikeImpactElectronCorrelations2024,oikeNonlinearMagnetoelectricEffect2024,mendozaNonlinearPhotomagnetizationInsulators2024}, which inevitably accompanies Joule heating, the light-induced NLEE is observable in insulators and is thus free from dissipation.
These advantages make the light-induced NLEE particularly suited for probing ferroaxial ordering, as most known ferroaxial materials~\cite{jinObservationFerrorotationalOrder2020,hayashidaPhaseTransitionDomain2021,hayashidaVisualizationFerroaxialDomains2020,yokotaThreedimensionalImagingFerroaxial2022,fangFerrorotationalSelectivityIlmenites2023} and candidates~\cite{johnsonCu3Nb2O8MultiferroicChiral2011,johnsonGiantImproperFerroelectricity2012,hanateFirstObservationSuperlattice2021,xuMultipleFerroicOrders2022,yangVisualizationChiralElectronic2022,jinVoltageEnablesFerrorotational2023,liuElectricalSwitchingFerrorotational2023,hayashidaElectricFieldInduced2023,kajitaFerroaxialTransitionsGlaseriteType2024} are insulating.
While second-harmonic generation (SHG) has been successfully used to image ferroaxial domains~\cite{yokotaThreedimensionalImagingFerroaxial2022}, a theoretical description beyond the electric-dipole approximation (including magnetic-dipole and/or electric-quadrupole terms) is generally required to capture axial-multipole contributions; in contrast, the present NLEE yields a static magnetization, offering a complementary, magnetization-based probe of ferroaxial order.

The rest of the paper is organized as follows.
Section~\ref{sec:formalism} discusses the relationship between the ET dipole and the NLEE tensor based on group-theoretical analysis and response theory.
By decomposing the polarization degrees of freedom of light, we clarify that both linearly polarized light (LPL) and circularly polarized light (CPL) couple to the ET dipole via distinct mechanisms.
To substantiate the symmetry-based predictions, we examine the NLEE under ferroaxial ordering using a minimal model in Sec.~\ref{sec:model_calculation}.
We show that the form of the SOC determines the direction of the induced spin magnetization; the ET dipole in Eq.~\eqref{eq:etd} tilts the spin-magnetization direction.
Section~\ref{sec:summary} summarizes the present paper and relates magnetic point groups (MPGs) to the NLEE.

\section{Formalism of the NLEE tensor}
\label{sec:formalism}

This section discusses the relationship between the NLEE and relevant multipoles.
We introduce the light-induced NLEE tensor as 
\begin{equation}
  \delta M_{\mu}^{\beta}=\frac{1}{2}\sum_{\nu\lambda}\sum_{\omega=\pm\Omega}\zeta_{\mu;\nu\lambda}^{\beta}(\omega,-\omega)E_{\nu}(\omega)E_{\lambda}(-\omega)
  \label{eq:nlee}
\end{equation}
for $\mu=x,y,z$.
The summations for $\nu$ and $\lambda$ runs over $x$, $y$, $z$.
The superscript $\beta=S$ and $L$ denotes spin and orbital contributions, respectively.
The tensor $\zeta_{\mu;\nu\lambda}(\Omega,-\Omega)$ is symmetric under the exchange of the last two subscripts and frequencies, i.e., $\zeta_{\mu;\nu\lambda}(\Omega,-\Omega)=\zeta_{\mu;\lambda\nu}(-\Omega,\Omega)$.
Since the electric field is real in the time domain, $E_{\nu}(\Omega)=E_{\nu}^{*}(-\Omega)$ holds in the frequency domain.
In the following, we omit $\beta$ unless otherwise specified.

\subsection{Polarization degrees of freedom of light}

We decompose the induced magnetization into contributions from the LPL and the CPL, analogous to standard photocurrent analysis~\cite{watanabeChiralPhotocurrentParityViolating2021}:
\begin{equation}
  \delta M_{\mu}=\delta M_{\mu}^{\mathrm{LP}}+\delta M_{\mu}^{\mathrm{CP}}.
  \label{eq:deltaM_decomposition}
\end{equation}
Each component is induced by the real and imaginary parts of the field product in Eq.~\eqref{eq:nlee}, 
\begin{subequations}
  \begin{equation}
    L_{\nu\lambda}(\Omega)=\mathrm{Re}[E_{\nu}(\Omega)E_{\lambda}^{*}(\Omega)],
    \label{eq:LP_field}
  \end{equation}
  \begin{equation}
    \mathbf{F}(\Omega)=-\frac{i}{2}\mathbf{E}(\Omega)\times\mathbf{E}^{*}(\Omega).
    \label{eq:CP_field}
  \end{equation}
  \label{eq:decomposition_electric_field}
\end{subequations}
Using $E_{\nu}(\Omega)E_{\lambda}^{*}(\Omega)=L_{\nu\lambda}(\Omega)+i\sum_{\rho}\epsilon_{\rho\nu\lambda}F_{\rho}(\Omega)$, the right-hand side of Eq.~\eqref{eq:deltaM_decomposition} is rewritten as 
\begin{subequations}
  \begin{equation}
    \delta M_{\mu}^{\mathrm{LP}}=\sum_{\nu\lambda}\eta_{\mu;\nu\lambda}(\Omega)L_{\mu\nu}(\Omega),
    \label{eq: M_LP}
  \end{equation}
  and
  \begin{equation}
    \delta M_{\mu}^{\mathrm{CP}}=\sum_{\rho}\xi_{\mu;\rho}(\Omega)F_{\rho}(\Omega).
    \label{eq: M_CP}
  \end{equation}
  \label{eq:deltaM_components}
\end{subequations}
Here,
\begin{subequations}
  \begin{equation}
    \eta_{\mu;\nu\lambda}(\Omega)\equiv\frac{1}{2}[\zeta_{\mu;\nu\lambda}(\Omega,-\Omega)+\zeta_{\mu;\lambda\nu}(\Omega,-\Omega)],
    \label{eq:LP_tensor}
  \end{equation}
  and
  \begin{equation}
    \xi_{\mu;\rho}(\Omega)\equiv i\sum_{\nu\lambda}\epsilon_{\rho\nu\lambda}\zeta_{\mu;\nu\lambda}(\Omega,-\Omega).
    \label{eq:CP_tensor}
  \end{equation}
  $\epsilon_{\rho\nu\lambda}$ is a totally antisymmetric Levi-Civita tensor.
  \label{eq:nlee_tensor}
\end{subequations}
Equation~\eqref{eq:LP_tensor}[\eqref{eq:CP_tensor}] represents the LPL (CPL)-induced NLEE tensor.
Using $\zeta_{\mu;\lambda\nu}(\Omega,-\Omega)=\zeta_{\mu;\nu\lambda}(-\Omega,\Omega)=\zeta_{\mu;\nu\lambda}^{*}(\Omega,-\Omega)$, we obtain $\eta_{\mu;\nu\lambda}(\Omega)=\mathrm{Re}[\zeta_{\mu;\nu\lambda}(\Omega,-\Omega)]$ and $\xi_{\mu;\rho}(\Omega)=-\sum_{\nu\lambda}\epsilon_{\rho\nu\lambda}\mathrm{Im}[\zeta_{\mu;\nu\lambda}(\Omega,-\Omega)]$.
Thus, the LPL (CPL) contribution corresponds to the real (imaginary) part of $\zeta_{\mu;\nu\lambda}(\Omega,-\Omega)$.

\subsection{Relevant multipoles of the NLEE tensor}

The decomposition in Eq.~\eqref{eq:deltaM_decomposition} with Eq.~\eqref{eq:deltaM_components} clarifies the correspondence between the NLEE tensor and relevant multipoles.
First, we focus on the LPL-induced NLEE tensor in Eq.~\eqref{eq:LP_tensor}.
Since the irradiating electric field in Eq.~\eqref{eq:LP_field} is a rank-2, TR-even polar tensor, the relevant multipoles are the even-parity rank-1, 2, and 3 multipoles~\cite{yatsushiroMultipoleClassification1222021}; the rank-1 magnetic or ET dipole $M_{1m}/G_{1m}$, the rank-2 electric or MT quadrupole $Q_{2m}/T_{2m}$, and the rank-3 magnetic or ET octupoles $M_{3m}/G_{3m}$.
For TR-symmetric systems, the detailed relationship between the LPL-induced NLEE tensor components and multipole degrees of freedom is given by
\begin{subequations}
  \label{eq:LP_tensor_MPs}
  \begin{equation}
      \underline{\eta}^{\mathrm{S}}=
      \begin{pmatrix}
        3G_{x}^{\prime}+2G_{x}^{\alpha} & G_{y}^{\prime}-G_{y}^{\alpha}-G_{y}^{\beta} & G_{z}^{\prime}-G_{z}^{\alpha}+G_{z}^{\beta}
        \\
        G_{x}^{\prime}-G_{x}^{\alpha}+G_{x}^{\beta} & 3G_{y}^{\prime}+2G_{y}^{\alpha} & G_{z}^{\prime}-G_{z}^{\alpha}-G_{z}^{\beta}
        \\
        G_{x}^{\prime}-G_{x}^{\alpha}-G_{x}^{\beta} & G_{y}^{\prime}-G_{y}^{\alpha}+G_{y}^{\beta} & 3G_{z}^{\prime}+2G_{z}^{\alpha}
        \\
        G_{xyz} & G_{z}^{\prime}-G_{z}^{\alpha}-G_{z}^{\beta} & G_{y}^{\prime}-G_{y}^{\alpha}+G_{y}^{\beta}
        \\
        G_{z}^{\prime}-G_{z}^{\alpha}+G_{z}^{\beta} & G_{xyz} & G_{x}^{\prime}-G_{x}^{\alpha}-G_{x}^{\beta}
        \\
        G_{y}^{\prime}-G_{y}^{\alpha}-G_{y}^{\beta} & G_{x}^{\prime}-G_{x}^{\alpha}+G_{x}^{\beta} & G_{xyz}
      \end{pmatrix}^{\mathrm{T}},
  \end{equation}
  \begin{equation}
      \underline{\eta}^{\mathrm{A}}=
      \begin{pmatrix}
        0 & 2(G_{y}-Q_{zx}) & 2(G_{z}+Q_{xy})
        \\
        2(G_{x}+Q_{yz}) & 0 & 2(G_{z}-Q_{xy})
        \\
        2(G_{x}-Q_{yz}) & 2(G_{y}+Q_{zx}) & 0 
        \\
        Q_{u}+Q_{v} & -(G_{z}-Q_{xy}) & -(G_{y}+Q_{zx})
        \\
        -(G_{z}+Q_{xy}) & -Q_{u}+Q_{v} & -(G_{x}-Q_{yz})
        \\
        -(G_{y}-Q_{zx}) & -(G_{x}+Q_{yz}) & -2Q_{v}
      \end{pmatrix}^{\mathrm{T}},
  \end{equation}
  with $\underline{\eta}=\underline{\eta}^{\mathrm{S}}+\underline{\eta}^{\mathrm{A}}$.
\end{subequations}
Here, we have decomposed $\eta_{\mu;\nu\lambda}$ into the totally symmetric component $\eta^{\mathrm{S}}_{\mu;\nu\lambda}$, which is symmetric for the permutations of subscripts $(\mu,\nu,\lambda)$, and the asymmetric component $\eta^{\mathrm{A}}_{\mu;\nu\lambda}$~\cite{tsirkinSeparationHallOhmic2022}.
$(G_{x},G_{y},G_{z})$ and $(G_{x}^{\prime},G_{y}^{\prime},G_{z}^{\prime})$ are the ET dipoles, $(Q_{u},Q_{v},Q_{yz},Q_{zx},Q_{xy})$ is the E quadrupole, and $(G_{xyz},G_{x}^{\alpha},G_{y}^{\alpha},G_{z}^{\alpha},G_{x}^{\beta},G_{y}^{\beta},G_{z}^{\beta})$ is the ET octupole.
We note that the same relationship can be obtained for the current-induced NLEE tensor defined by the static limit 
\begin{equation*}
  \eta_{\mu;\nu\lambda}\equiv\lim_{\omega_{1},\omega_{2}\to 0}\eta_{\mu;\nu\lambda}(\omega_{1},\omega_{2}).
\end{equation*}

By contrast, the electric field in Eq.~\eqref{eq:CP_field} transforms as an ET dipole, i.e., a rank-1, TR-even axial tensor.
The relevant multipoles of the CPL-induced NLEE tensor in Eq.~\eqref{eq:CP_tensor} are the even-parity rank-0, 1, and 2 ones; the rank-0 electric or MT monopole $Q_{0}/T_{0}$, the rank-1 magnetic or ET dipole $M_{1m}/G_{1m}$, and the rank-2 magnetic or MT quadrupole $Q_{2m}/T_{2m}$.
The correspondence between the components of the CPL-induced NLEE tensor and multipole degrees of freedom in the TR-symmetric system is given by 
\begin{equation}
  \underline{\xi}=
  \begin{pmatrix}
    Q_{0}-Q_{u}+Q_{v} & Q_{xy}+G_{z} & Q_{zx}-G_{y}
    \\
    Q_{xy}-G_{z} & Q_{0}-Q_{u}-Q_{v} & Q_{yz}+G_{x}
    \\
    Q_{zx}+G_{y} & Q_{yz}-G_{x} & Q_{0}+2Q_{u}
  \end{pmatrix}.
  \label{eq:CP_tensor_MPs}
\end{equation}
The additional monopole contribution characterizes the antisymmetric part of a second-order nonlinear external field~\cite{kandaNonlinearFrequencyasymmetricOptical2025}.
The contributions of the magnetic-type multipoles are obtained by exchanging $(Q,G)\to (T,M)$ in Eqs.~\eqref{eq:LP_tensor_MPs} and \eqref{eq:CP_tensor_MPs}.

We summarize the relationship between the polarization degrees of freedom of light and the relevant multipoles in Table~\ref{tab:relevant_MP}.
\begin{table}[t]
  \centering
  \caption{Relationship between the nonlinear Edelstein effect induced by the linearly polarized light (LPL)/circularly polarized light (CPL) and relevant multipoles denoted by $\checkmark$.}
  \begin{tabular}{ll|cc}
    \hline\hline
    Multipoles & & LPL & CPL
    \\
    \hline
    Electric/MT monopole & $Q_{0}/T_{0}$ & --- & $\checkmark$
    \\
    Magnetic/ET dipole & $M_{1m}/G_{1m}$ & $\checkmark$ & $\checkmark$ 
    \\
    Electric/MT quadrupole & $Q_{2m}/T_{2m}$ & $\checkmark$ & $\checkmark$
    \\
    Magnetic/ET octupole & $M_{3m}/G_{3m}$ & $\checkmark$ & ---
    \\
    \hline\hline
  \end{tabular}
  \label{tab:relevant_MP}
\end{table}
We find that both the LPL and CPL can detect ferroaxial ordering expressed by the ET dipole $G_{1m}$.
Meanwhile, we shall see that the transition processes are different between them.

\subsection{Microscopic transition processes}

Next, we discuss the transition processes of the light-induced NLEE.
We derive the NLEE tensor within the length gauge using the quantum kinetic theory of the density matrix operator~\cite{oikeNonlinearMagnetoelectricEffect2024,watanabeChiralPhotocurrentParityViolating2021}.
The details of the derivation are given in Appendix~\ref{sec:derivation}.
The nonlinear response tensor in Eq.~\eqref{eq:nlee} is decomposed as 
\begin{equation*}
  \zeta=\zeta^{(\mathrm{JJ})}+\zeta^{(\mathrm{EJ})}+\zeta^{(\mathrm{JE})}+\zeta^{(\mathrm{EE})},
\end{equation*} 
where the superscript $\mathrm{J}$$(\mathrm{E})$ denotes the intraband (interband) processes.
\begin{table}[t]
  \centering
  \caption{Relationship between polarization degrees of freedom of light and transition processes.
  $(\mathrm{P})$ and $(\delta)$ in the first column stand for the reactive part and absorptive part.
  The superscript $\mathrm{d}$ $(\mathrm{o})$ means the intraband (interband) part of the Bloch representation of the angular momentum operator.
  The second column shows the presence $\checkmark$ of the transition processes in insulators.
  The third (fourth) column indicates contributions coming from the even-parity electric or ET (magnetic or MT) multipoles.
  $\updownarrow$ and $\circlearrowright$ are the linearly polarized light and circularly polarized light, respectively.}
  \begin{tabular}{l|c|cc}
    \hline\hline
    \begin{tabular}{c}
      \\
      Process
      \\
    \end{tabular} & 
    \begin{tabular}{c}
      \\
      Insulator
      \\
    \end{tabular} & 
    \begin{tabular}{c}
      TR-even
      \\
      (electric or ET)
    \end{tabular} & 
    \begin{tabular}{c}
      TR-odd 
      \\
      (magnetic or MT)
    \end{tabular}
    \\
    \hline
    $\zeta^{(\mathrm{JJ})}$ & --- & --- & $\updownarrow$
    \\
    $\zeta^{(\mathrm{EJ})}$ & --- & $\circlearrowright$ & ---
    \\
    $\zeta^{(\mathrm{EE;d})}(\delta)$ & $\checkmark$ & $\circlearrowright$ & $\updownarrow$
    \\
    $[\zeta^{(\mathrm{JE})}+\zeta^{(\mathrm{EE;o})}](\delta)$ & $\checkmark$ & $\updownarrow$ & $\circlearrowright$ 
    \\
    $[\zeta^{(\mathrm{JE})}+\zeta^{(\mathrm{EE;o})}](\mathrm{P})$ & $\checkmark$ & $\circlearrowright$ & $\updownarrow$ 
    \\
    \hline\hline
  \end{tabular}
  \label{tab:process_NLEE}
\end{table}
Table~\ref{tab:process_NLEE} shows the relationship between the polarization degrees of freedom of light and the transition processes.
The TR-odd terms contribute when both TR and $\mathcal{PT}$(product of the SI and TR) symmetries are broken, where the relevant multipoles are odd-rank magnetic multipoles or even-rank MT multipoles.

Focusing on TR-symmetric insulators, we find that the LPL-induced NLEE requires resonance, whereas the CPL-induced NLEE tensor includes both reactive and absorptive parts.
For simplicity, we consider the optical regime $\Omega\tau\gg 1$ for the light frequency $\Omega$ and the relaxation time $\tau$, where the absorptive part dominates.

\begin{subequations}
  In such a situation, the LPL-induced tensor reads
  \begin{equation}
    \eta_{\mu;\nu\lambda}(\Omega)=
    \frac{\pi\mu_{\mathrm{B}}e^{2}}{2N}\sum_{\mathbf{k}ab}\delta(\hbar\Omega-\varepsilon_{ab})\mathrm{Re}\left(Q_{ab}^{\mu;\nu\lambda}+Q_{ab}^{\mu;\lambda\nu}\right)f_{ab},
    \label{eq:lp_nlee_insulator}
  \end{equation}
  where standard notation follows Appendix~\ref{sec:derivation}.
  We have omitted the $\mathbf{k}$ dependence of the quantities in the summation and divided the tensor by the number of the unit cell $N$.
  The LPL-induced NLEE is independent of the relaxation time and essentially three-band process.

  Meanwhile, the CPL-induced tensor is
  \begin{equation}
    \xi_{\mu;\rho}(\Omega)=\frac{\pi\mu_{\mathrm{B}}e^{2}\tau}{2\hbar N}\sum_{\mathbf{k}ab}\delta(\hbar\Omega-\varepsilon_{ab})
    \sum_{\nu\lambda}\epsilon_{\rho\nu\lambda}\mathrm{Im}\left([\beta^{\mu},A^{\nu}]_{ab}^{\prime}A_{ba}^{\lambda}\right)f_{ab},
    \label{eq:cp_nlee_insulator}
  \end{equation}
  where $A^{\nu}_{ab}$ is the interband Berry connection.
  Equation~\eqref{eq:cp_nlee_insulator}, which is the extension of Eq.~(4) in Ref.~\cite{xuLightinducedStaticMagnetization2021} to the case of the band degeneracy due to $\mathcal{PT}$ symmetries, represents the contribution from the spin $(\beta_{\mu}=2s_{\mu})$ or orbital $(\beta_{\mu}=l_{\mu})$ polarization difference between conduction and valence bands.
  \label{eq:nlee_insulator}
\end{subequations}
The CPL-induced NLEE tensor is proportional to the relaxation time $\tau$, which is a characteristic of the optical spin injection~\cite{oestreichSpinInjectionSpin2002,zuticSpintronicsFundamentalsApplications2004,nastosFullBandStructure2007,xuLightinducedStaticMagnetization2021,fregosoBulkPhotospinEffect2022,oikeImpactElectronCorrelations2024}.
To examine spin injection in more detail, we assume that there is no band degeneracy and set $\rho=z$.
Introducing the interband Berry connection for the circular polarization by $A_{ab}^{z(\pm)}=(A_{ab}^{x}\pm iA_{ab}^{y})/\sqrt{2}$~\cite{watanabeChiralPhotocurrentParityViolating2021}, we can rewrite Eq.~\eqref{eq:cp_nlee_insulator} as follows:
\begin{equation}
  \xi_{\mu;z}(\Omega)=
  \frac{\pi\mu_{\mathrm{B}}e^{2}\tau}{2\hbar N}\sum_{\mathbf{k}ab}\delta(\hbar\Omega-\varepsilon_{ab})\Delta^{(h)\mu}_{ab}\left(|A^{z(+)}_{ab}|^{2}-|A^{z(-)}_{ab}|^{2}\right)f_{ab},
\end{equation}
where $\Delta^{(h)\mu}_{ab}=\beta^{\mu}_{aa}-\beta^{\mu}_{bb}$ is the magnetization difference of the two bands and we have used the relation $A_{ba}^{z(\mp)}=(A_{ab}^{z(\pm)})^{*}$.
This formula explicitly indicates that nonzero optical orbital/spin injection requires orbital/spin polarization differences between bands and antisymmetric transition between left/right CPLs.

\section{Model calculation}
\label{sec:model_calculation}

To investigate the validity of symmetry and microscopic transition process analyses, we evaluate the light-induced NLEE under ferroaxial ordering based on a minimal model.

\subsection{Hamiltonian}

Here, we consider the three $p$ orbitals $(\phi_{x},\phi_{y},\phi_{z})$ on a triangular lattice stacked along the $z$ axis, whose point group is $D_{6h}$.
The lattice vectors are $\mathbf{a}_{1}=a(1,0,0)$, $\mathbf{a}_{2}=a(-1/2,\sqrt{3}/2,0)$, and $\mathbf{a}_{3}=c(0,0,1)$ with lattice constants $a$ and $c$.
The tight-binding Hamiltonian is given by 
\begin{equation}
  H_{0}=\sum_{\mathbf{k}}\sum_{\alpha\beta}h_{\alpha\beta}(\mathbf{k})c_{\mathbf{k}\alpha}^{\dag}c_{\mathbf{k}\beta},
\end{equation}
where $c_{\mathbf{k}\alpha}^{\dag}$ $(c_{\mathbf{k}\alpha})$ is a creation (an annihilation) operator of an electron with the crystal momentum $\mathbf{k}$ and a set of the internal degrees of freedom $\alpha$, which includes the orbital $l\in\{x,y,z\}$ and spin $\sigma\in\{\uparrow,\downarrow\}$. 
$h_{\alpha\beta}(\mathbf{k})$ corresponds to the Hamiltonian matrix denoted by $\hat{h}(\mathbf{k})$, which consists of four distinct parts:
\begin{equation}
  \hat{h}(\mathbf{k})=\hat{h}_{\mathrm{hop}}(\mathbf{k})+\hat{h}_{\mathrm{CEF}}+\lambda\mathbf{l}\cdot\mathbf{s}+g_{z}(\mathbf{l}\times\mathbf{s})_{z}.
  \label{eq:Hamiltonian_matrix}
\end{equation}
The first term, $\hat{h}_{\mathrm{hop}}(\mathbf{k})$, is the hopping term.
We consider the nearest-neighbor (NN) intralayer hopping in the $\pm\mathbf{a}_{1}$, $\pm\mathbf{a}_{2}$, and $\pm(\mathbf{a}_{1}+\mathbf{a}_{2})$ directions with amplitudes $(t_{1\sigma},t_{1\pi})$; the NN interlayer hopping in the $\pm\mathbf{a}_{3}$ directions with amplitudes $(t_{c\sigma},t_{c\pi})$; and the interlayer diagonal hopping in the  $\pm(\mathbf{a}_{1}\pm\mathbf{a}_{3})$, $\pm(\mathbf{a}_{2}\pm\mathbf{a}_{3})$, $\pm(\mathbf{a}_{1}+\mathbf{a}_{2}\pm\mathbf{a}_{3})$ directions with amplitudes $(t_{2\sigma},t_{2\pi})$.
Finally, the hopping Hamiltonian is expressible by using the multipole basis set as follows~\cite{kusunoseSymmetryadaptedModelingMolecules2023}:
\begin{equation}
  \begin{aligned}
    \hat{h}_{\mathrm{hop}}(\mathbf{k})=&\, f_{0}(\mathbf{k})\mathbb{Q}_{0}+f_{u}(\mathbf{k})\mathbb{Q}_{u}+f_{v}(\mathbf{k})\mathbb{Q}_{v}
    \\
    &\, +f_{xy}(\mathbf{k})\mathbb{Q}_{xy}+f_{yz}(\mathbf{k})\mathbb{Q}_{yz}+f_{zx}(\mathbf{k})\mathbb{Q}_{zx},
  \end{aligned}
\end{equation}
where $\mathbb{Q}_{i}$ for $i=0$, $u$, $v$, $yz$, $zx$, $xy$ are the multipole bases in orbital space whose matrix elements are given by Eq.~\eqref{eq:mp_basis_orbital}, and 
\begin{equation}
  \begin{aligned}
    f_{0}(\mathbf{k})=&\, -\sqrt{2}\bar{t}_{1}Q_{0}(\mathbf{k})-\frac{2}{\sqrt{6}}\bar{t}_{c}Q_{0}^{(c)}(\mathbf{k})-2\bar{t}_{2}Q_{0}(\mathbf{k})Q_{0}^{(c)}(\mathbf{k}),
    \\
    f_{u}(\mathbf{k})=&\, \tilde{t}_{1}Q_{0}(\mathbf{k})-\frac{2}{\sqrt{3}}\tilde{t}_{c}Q_{0}^{(c)}(\mathbf{k})-\sqrt{2}\frac{2c^{2}-a^{2}}{a^{2}+c^{2}}\tilde{t}_{2}Q_{0}(\mathbf{k})Q_{0}^{(c)}(\mathbf{k}),
    \\
    f_{\mu}(\mathbf{k})=&\, \frac{\sqrt{6}}{2}\tilde{t}_{1}Q_{\mu}(\mathbf{k})+\sqrt{3}\frac{a^{2}}{a^{2}+c^{2}}\tilde{t}_{2}Q_{\mu}(\mathbf{k})Q_{0}^{(c)}(\mathbf{k});\ (\mu=xy,v),
    \\
    f_{\tau}(\mathbf{k})=&\, 2\sqrt{3}\frac{ac}{a^{2}+c^{2}}\tilde{t}_{2}Q_{\tau}(\mathbf{k});\ (\tau=yz,zx),
  \end{aligned}
\end{equation}
with the form factors 
\begin{equation}
  \begin{aligned}
    Q_{0}(\mathbf{k})=&\, \frac{\sqrt{6}}{3}\left[\cos{(k_{x}a)}+2\cos{\left(\frac{k_{x}a}{2} \right)}\cos{\left(\frac{\sqrt{3}k_{y}a}{2} \right)}\right],
    \\
    Q_{0}^{(c)}(\mathbf{k})=&\, \sqrt{2}\cos{(k_{z}c)},
    \\
    Q_{v}(\mathbf{k})=&\, -\frac{2\sqrt{3}}{3}\left[\cos{(k_{x}a)}-\cos{\left(\frac{k_{x}a}{2}\right) }\cos{\left(\frac{\sqrt{3}k_{y}a}{2}\right)}\right],
    \\
    Q_{xy}(\mathbf{k})=&\, 2\sin{\left(\frac{k_{x}a}{2} \right)}\sin{\left(\frac{\sqrt{3}k_{y}a}{2} \right)},
    \\
    Q_{yz}(\mathbf{k})=&\, 2\sqrt{2}\cos{\left(\frac{k_{x}a}{2} \right)}\sin{\left(\frac{\sqrt{3}k_{y}a}{2} \right)}\sin{(k_{z}c)},
    \\
    Q_{zx}(\mathbf{k})=&\, \frac{2\sqrt{6}}{3}\sin{(k_{z}c)}\left[\sin{(k_{x}a)}+\sin{\left(\frac{k_{x}a}{2} \right)}\cos{\left(\frac{\sqrt{3}k_{y}a}{2} \right)} \right].
  \end{aligned}
\end{equation}
We have also introduced $\bar{t}_{\nu}=t_{\nu\sigma}+2t_{\nu\pi}$ and $\tilde{t}_{\nu}=t_{\nu\sigma}-t_{\nu\pi}$ for $\nu=1$, $2$, and $c$.
The second term in Eq.~\eqref{eq:Hamiltonian_matrix}, $\hat{h}_{\mathrm{CEF}}$, is the crystal electric field (CEF).
Since $p$ orbitals under $D_{\mathrm{6h}}$ split into a nondegenerate $\phi_{z}$ state and two degenerate $(\phi_{x},\phi_{y})$ states, the CEF Hamiltonian is parametrized as 
\begin{equation}
  \hat{h}_{\mathrm{CEF}}=\Delta\mathbb{Q}_{u},
\end{equation}
where $\Delta>0$ places the $(\phi_{x},\phi_{y})$ orbitals in the ground state.
The third and fourth terms in Eq.~\eqref{eq:Hamiltonian_matrix} are the relativistic and the ferroaxial-origin SOC, respectively.
The latter is microscopically derived by the effective coupling between vertical-mirror-symmetry-breaking orbital hybridization and relativistic SOC~\cite{indaEmergentCrossproducttypeSpinorbit2025}.
Here, we take the $z$ component of the ET dipole $G_{z}$ as the order parameter;
when $g_{z}\neq 0$, the vertical mirror symmetry is broken and the symmetry is lowered to $C_{\mathrm{6h}}$.

\begin{figure}[t]
  \centering
  \includegraphics[width=\linewidth]{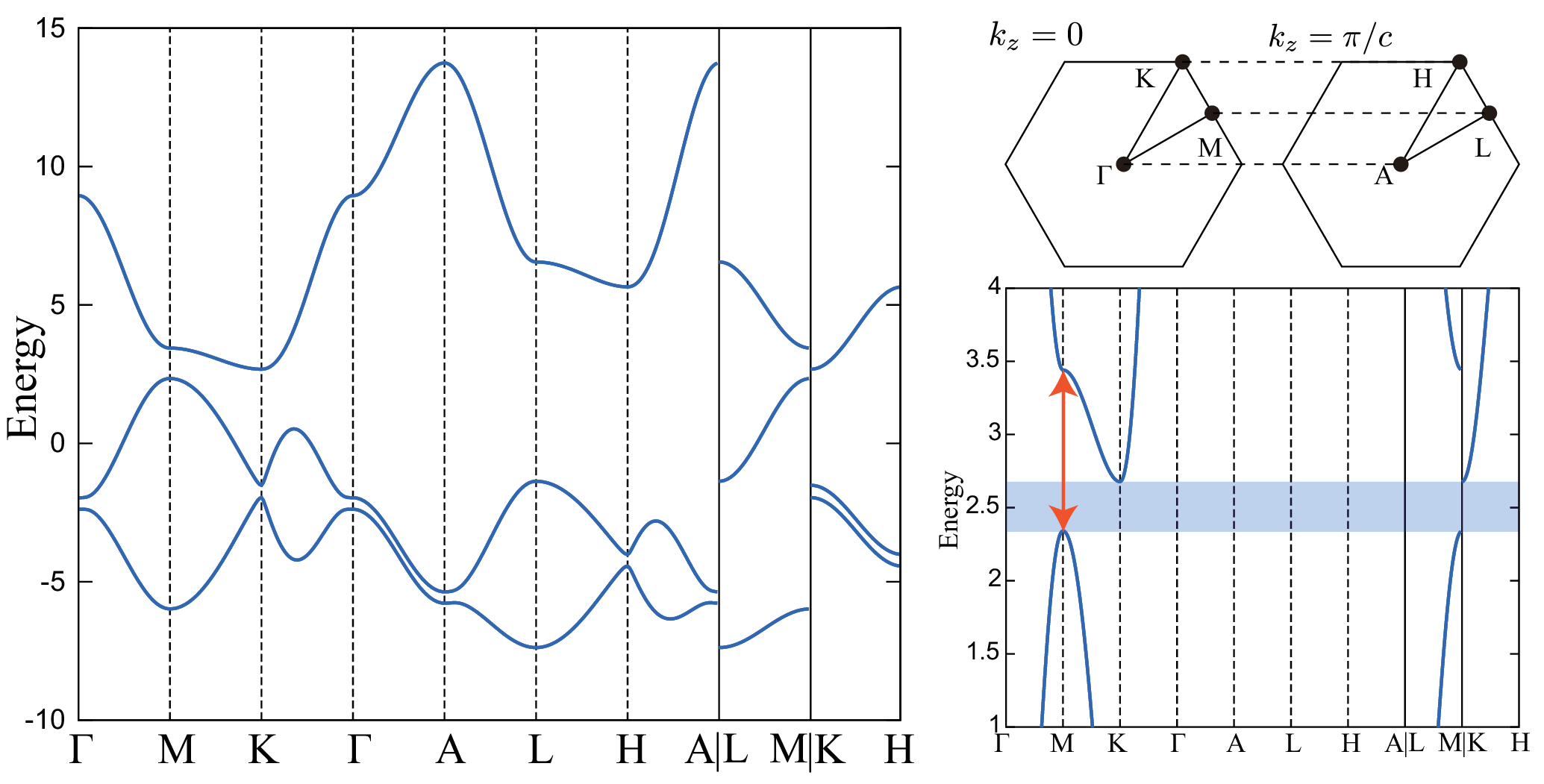}
  \caption{Band structure calculated by using parameters in Eq.~\eqref{eq:parameters} along the high symmetry lines in the Brillouin zone as shown on the right-top panel.
  The right-bottom figure shows the band gap between the highest and middle bands.
  The blue shaded area is the global band gap that realizes the band insulator, whereas the red arrow at the M point indicates the minimum local band gap that sets the absorption condition for the NLEE.}
  \label{fig:band_structure}
\end{figure}

Figure~\ref{fig:band_structure} shows the band structure with the following parameters:
\begin{equation}
  \begin{aligned}
    &\, \bar{t}_{1}=-0.6, \tilde{t}_{1}=1.8, \bar{t}_{2}=0.0, \tilde{t}_{2}=0.3, 
    \\
    &\, \bar{t}_{c}=-0.5, \tilde{t}_{c}=1.6,
    \\
    &\, \Delta=8, \lambda=0.4, g_{z}=0.5, a=c=1.
  \end{aligned}
  \label{eq:parameters}
\end{equation}
To achieve a band-insulating system, we adopt a sufficiently large $\Delta$.
All bands are doubly degenerate owing to $\mathcal{PT}$ symmetry.
We distinguish two types of band gaps.
One is the global band gap with the magnitude of $0.34$ (blue shaded area in Fig.~\ref{fig:band_structure}), which determines whether the system is insulating or metallic.
In the following, we evaluate the NLEE under the condition that the chemical potential $\mu$ is fixed within the global band gap, thereby maintaining the insulating state with a fixed filling of $n=4$.
In our parameter set, valence and conduction bands are dominated by $(\phi_{x},\phi_{y})$ and $\phi_{z}$, respectively.
Another is the direct band gap with the magnitude of $1.1$ (red arrow in Fig.~\ref{fig:band_structure}), which gives the absorption condition for the NLEE.

\subsection{Results}

\begin{figure*}[htbp]
  \centering
  \includegraphics[width=\linewidth]{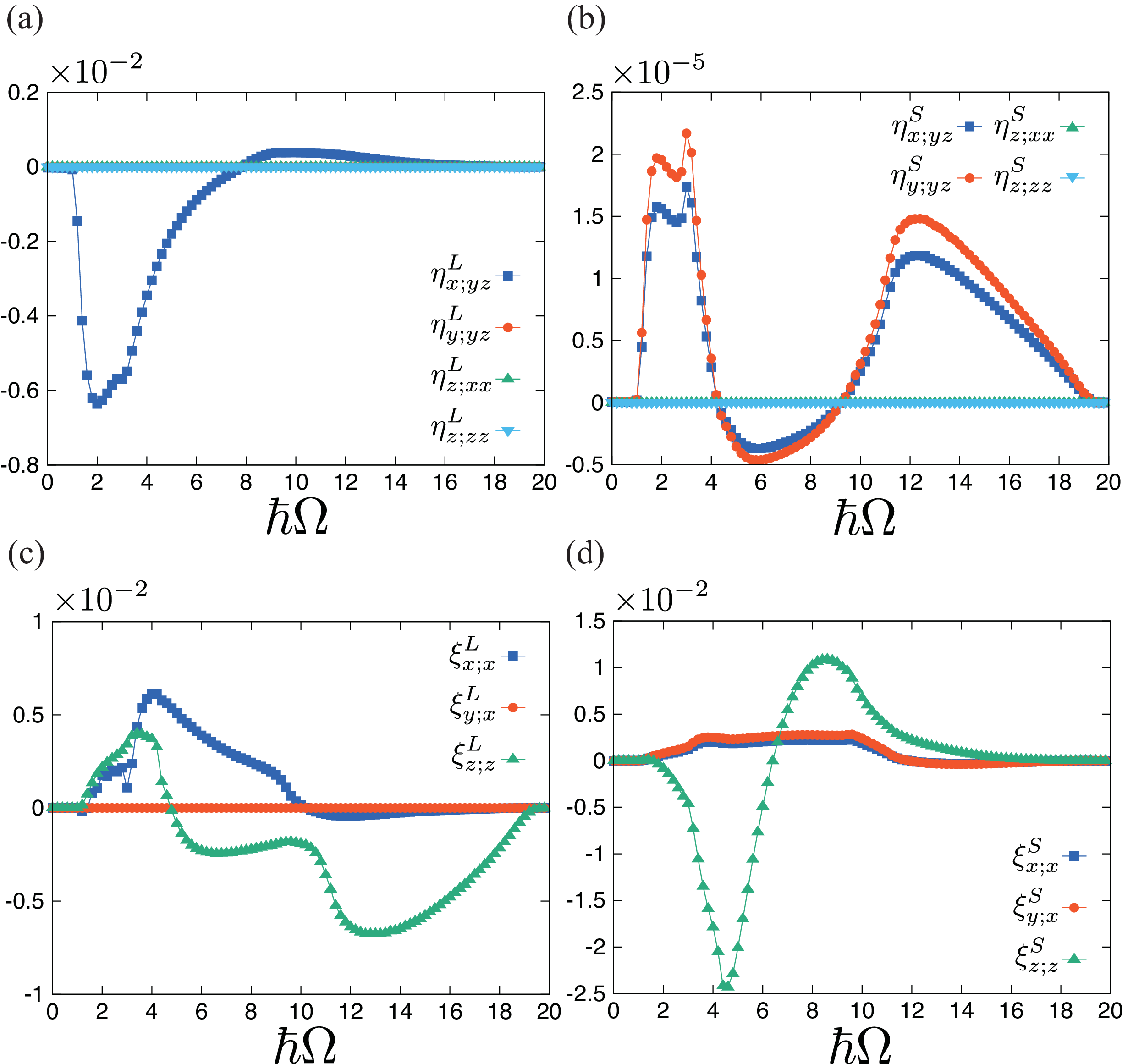}
  \caption{The calculated NLEE tensors (a) $\eta_{\mu;\nu\lambda}^{L}$, (b) $\eta_{\mu;\nu\lambda}^{S}$, (c) $\xi_{\mu;\rho}^{L}$, and (d) $\xi_{\mu;\rho}^{S}$.
  Here, the superscripts $L$ and $S$ represent the orbital- and spin-magnetization components, respectively.
  We set $e=\mu_{\mathrm{B}}=\hbar=1$, the scattering rate $\gamma=10^{-2}$, the temperature $T=10^{-3}$, and the number of the unit cell $N=768\times768\times512$, respectively.}
  \label{fig:nlee_frequency}
\end{figure*}

Figure~\ref{fig:nlee_frequency} shows the frequency dependence of the symmetry-allowed NLEE components evaluated by Eq.~\eqref{eq:nlee_insulator}.
In the normal state ($g_{z}=0$, point group $D_{6h}$), the allowed components are $\eta_{x;yz}=-\eta_{y;zx}$ (LPL) and $\xi_{x;x}=\xi_{y;y}$, $\xi_{z;z}$ (CPL).
In the ferroaxial state ($g_{z} \neq 0$, $C_{6h}$), additional components would appear:
$\eta_{z;xx}=\eta_{z;yy}$, $\eta_{z;zz}$, $\eta_{x;zx}=\eta_{y;yz}$ (LPL) and $\xi_{x;y}=-\xi_{y;x}$ (CPL).
As an overall tendency in Fig.~\ref{fig:nlee_frequency}, the NLEE becomes finite when the photon energy exceeds the minimum local band gap energy, enabling optical transitions.
The reactive part neglected in Eq.~(\ref{eq:cp_nlee_insulator}) and the relaxation time dependences are examined in Appendix~\ref{sec:relaxation_time}.

First, we discuss the LPL-induced NLEE tensor, as shown in Figs.~\ref{fig:nlee_frequency}(a) and~\ref{fig:nlee_frequency}(b).
Based on the parameters in Eq.~\eqref{eq:parameters}, the tensor components $\eta_{x;yz}^{L}$, $\eta_{x;yz}^{S}$, and $\eta_{y;yz}^{S}$ become nonzero, while the others are zero.
Among them, $\eta_{x;yz}^{L}$ and $\eta_{x;yz}^{S}$ remain finite even in the normal state with $g_{z}=0$, whereas $\eta_{y;yz}^{S}$ vanishes as $g_{z} \to 0$.
These results are consistent with the above symmetry analysis.
To clarify the microscopic origin of the NLEE, we investigate the essential model parameters~\cite{oiwaSystematicAnalysisMethod2022}.
In the case of the orbital-magnetization component, i.e., $\delta M_{x}^{L}$, we find that the following two factors are necessary:
One is the orbital hybridization between $\phi_{y}$ and $\phi_{z}$, which means the appearance of the electric quadrupole degree of freedom $Q_{yz}$.
The other is the different orbital occupation between the $\phi_{y}$ and $\phi_{z}$, induced by hopping processes involving other electric quadrupole degrees of freedom $Q_{u}$ or $Q_{v}$ (see the matrix elements in Appendix~\ref{sec:multipoles_p_orbital}).
Since $Q_{yz}$ arises when $\tilde{t}_{2}\neq0$, $\tilde{t}_{2}\neq0$ corresponds to the essential parameter for $\eta_{x;yz}^{L}$.
This conclusion is supported by numerical calculations, which confirm that all tensor components vanish by setting $\tilde{t}_{2}=0$.
Meanwhile, the relativistic SOC plays a less significant role in $\delta M_{x}^{L}$; $\eta_{x;yz}^{L}$ remains nonzero for $\lambda=0$.

In contrast with the orbital-magnetization component, the two types of SOCs play an essential role in inducing the spin-magnetization components.
This tendency is numerically confirmed; $\eta_{x;yz}^{S}$ ($\eta_{y;yz}^{S}$) vanishes for $\lambda=0$ ($g_{z}=0$).
This suggests that the onset of spin magnetization is interpreted as a conversion of orbital magnetization by the emergent SOC as follows:
Let us consider the case of the tensor component $\eta_{x;yz}$, where the $x$ component of the magnetization induced by the LPL as $L_{yz}(\Omega)=\mathrm{Re}[E_{y}(\Omega)E_{z}^{*}(\Omega)]$.
As discussed above, $L_{yz}(\Omega)$ generates the orbital magnetization $\delta M_{x}^{L}=\eta_{x;yz}^{L}(\Omega)L_{yz}(\Omega)$ without the SOC.
In such a situation, by combining the effect of the relativistic SOC, the LPL generates the spin magnetization in the same direction as the orbital magnetization, i.e., $\delta M_{x}^{S}=\eta_{x;yz}^{S}(\Omega)L_{yz}(\Omega)$, since the relativistic SOC couples the same component of the orbital and spin angular momenta via the scalar product, $\mathbf{l}\cdot\mathbf{s}$.
Similarly, one can interpret the emergence of $\eta_{y;yz}^{S}$ under ferroaxial ordering via the effective SOC with the cross-product form $\mathbf{l}\times \mathbf{s}$. 
In other words, owing to the coupling as $l_{x}s_{y}-l_{y}s_{x}$, the LPL $L_{yz}(\Omega)$ leads to the $y$ component of the spin magnetization, $\delta M_{y}^{S}=\eta_{y;yz}^{S}(\Omega)L_{yz}(\Omega)$, under ferroaxial ordering.
This result indicates that the direction of the induced magnetization is related to the form of the SOC, as detailed in Sec.~\ref{sec:discussion}.

Next, we discuss the CPL-induced NLEE tensor, as shown Figs.~\ref{fig:nlee_frequency}(c) and~\ref{fig:nlee_frequency}(d).
Among the symmetry-allowed tensor components, $\xi_{x;x}^{L}$, $\xi_{z;z}^{L}$, $\xi_{x;x}^{S}$, $\xi_{z;z}^{S}$, and $\xi_{y;x}^{S}$ are nonzero in the present model parameters in Eq.~\eqref{eq:parameters}.
In particular, $\xi_{y;x}^{S}$ arises only under ferroaxial ordering.
The essential parameters for the CPL-induced NLEE are almost the same as those for the LPL-induced NLEE.
For the in-plane orbital-magnetiztion component, $\xi_{x;x}^{L}$, it emerges due to the hybridization and polarization in orbital space, which is similar to the case of $\eta_{x;yz}^{L}$.
The former is related to the activation of $Q_{yz}$ hybridizing $\phi_{y}$ and $\phi_{z}$, and the latter is related to the activation of $(Q_{u},Q_{v})$.
Thus, the hopping parameter $\tilde{t}_{2}$ is essential. 
On the other hand, in contrast with the LPL-induced NLEE, SOC is also needed for $\xi_{x;x}^{L}$ because of the requirement of the orbital hybridization for $l^{x}_{ac}\neq 0$ ($\varepsilon_{a}=\varepsilon_{c}$)~\footnote{SOC is not essential in orbital magnetization in general.
For instance, as an effect of polarization along the $x$ axis, we add the following hopping term to the Hamiltonian:
$\hat{h}_{\mathrm{hop}}^{\prime}(\mathbf{k})=t^{\prime}\sin{(k_{z}c)}\mathbb{M}_{y}$.
When $t^{\prime}\neq0$, we obtain $\xi_{x;x}^{L}\neq0$, that is, orbital magnetization is driven even without SOC.}.
For the in-plane spin-magnetization components, $\xi_{x;x}^{S}$, and $\xi_{y;x}^{S}$, their emergence is understood by the coupling between $\xi_{x;x}^{L}$ and two types of SOCs;
the relativistic SOC gives the spin magnetization parallel to the orbital magnetization, $\xi_{x;x}^{S}$, whereas the ferroaxial-origin SOC contributes to the perpendicular component, $\xi_{y;x}^{S}$.
Thus, like the LPL, the orbital magnetization is converted to the spin magnetization via the SOC.
For the out-of-plane orbital-magnetization component, $\xi_{z;z}^{L}$, we numerically find that $\tilde{t}_{2}$ is not necessary in contrast with the in-plane component to induce $\xi_{z;z}^{L}$ because the polarization between $\phi_{x}$ and $\phi_{y}$ already occurred by the intralayer hopping.
For the out-of-plane spin-magnetization component, $\xi_{z;z}^{S}$, it also becomes nonzero even for $\lambda=0$ once the ferroaxial moment is present, i.e., $g_{z}\neq 0$.
This is attributed to the fact that the second- or higher-order coupling of the ferroaxial-origin SOC leads to a term proportional to $\sigma_{z}$.

\subsection{Discussion}
\label{sec:discussion}

Finally, we discuss the SOC dependence for the NLEE.
To this end, we set the direction of the incident light as 
\begin{equation}
  \mathbf{e}_{\mathrm{inc}}=(\sin{\theta}\cos{\phi}, \sin{\theta}\sin{\phi}, \cos{\theta}),
\end{equation}
where $0\leq\theta<\pi$ and $0\leq\phi<2\pi$, and two degrees of freedom of the transverse light as 
\begin{equation}
  \mathbf{e}_{1}=(-\sin{\phi}, \cos{\phi}, 0),
  \mathbf{e}_{2}=(-\cos{\theta}\cos{\phi}, -\cos{\theta}\sin{\phi}, \sin{\theta}).
\end{equation}
Here, $\mathbf{e}_{\mathrm{inc}}$, $\mathbf{e}_{1}$, and $\mathbf{e}_{2}$ are orthornormal.
Using these unit vectors, the LPL and CPL can be described as $\mathbf{e}_{\mathrm{LP}}=\cos{\alpha}\mathbf{e}_{1}+\sin{\alpha}\mathbf{e}_{2}$, and $\mathbf{e}_{\mathrm{CP}}^{\pm}=(\mathbf{e}_{1}\pm i\mathbf{e}_{2})/\sqrt{2}$, where $\alpha$ ($0\leq\alpha<\pi$) represents the angle of the LPL measured from the $\mathbf{e}_{1}$ direction.
To see the induced magnetization under ferroaxial ordering along the $z$ axis, we set $\theta=\pi/2$, that is, the light propagates in the $xy$-plane.
\begin{subequations}
  We consider the components of magnetization parallel and perpendicular to the direction of the light:
  \begin{equation}
    \delta M_{\parallel}\equiv\delta\mathbf{M}\cdot\mathbf{e}_{\mathrm{inc}}
    =
    \begin{cases}
      \sin{(2\alpha)}\eta_{x;yz}|E|^{2} & (\mathrm{LP})
      \\
      \mp\frac{1}{2}\xi_{x;x}|E|^{2} & (\mathrm{CP})
    \end{cases},
  \end{equation}
  \begin{equation}
    \delta M_{\perp}\equiv\delta\mathbf{M}\cdot\mathbf{e}_{1}
    =
    \begin{cases}
      \sin{(2\alpha)}\eta_{y;yz}|E|^{2} & (\mathrm{LP})
      \\
      \mp\frac{1}{2}\xi_{y;x}|E|^{2} & (\mathrm{CP})
    \end{cases}.
  \end{equation}
  We have omitted the frequency dependence and assumed $\alpha\neq 0,\pi/2$ for the LPL, that is, the electric field has both $xy$-plane and $z$ components.
  \label{eq:magnetization_component}
\end{subequations}
Since Eq.~\eqref{eq:magnetization_component} is independent of the direction of the light, we assume $\mathbf{e}_{\mathrm{inc}}=(1,0,0)$ for simplicity.

\begin{figure*}[htbp]
  \centering
  \includegraphics[width=\linewidth]{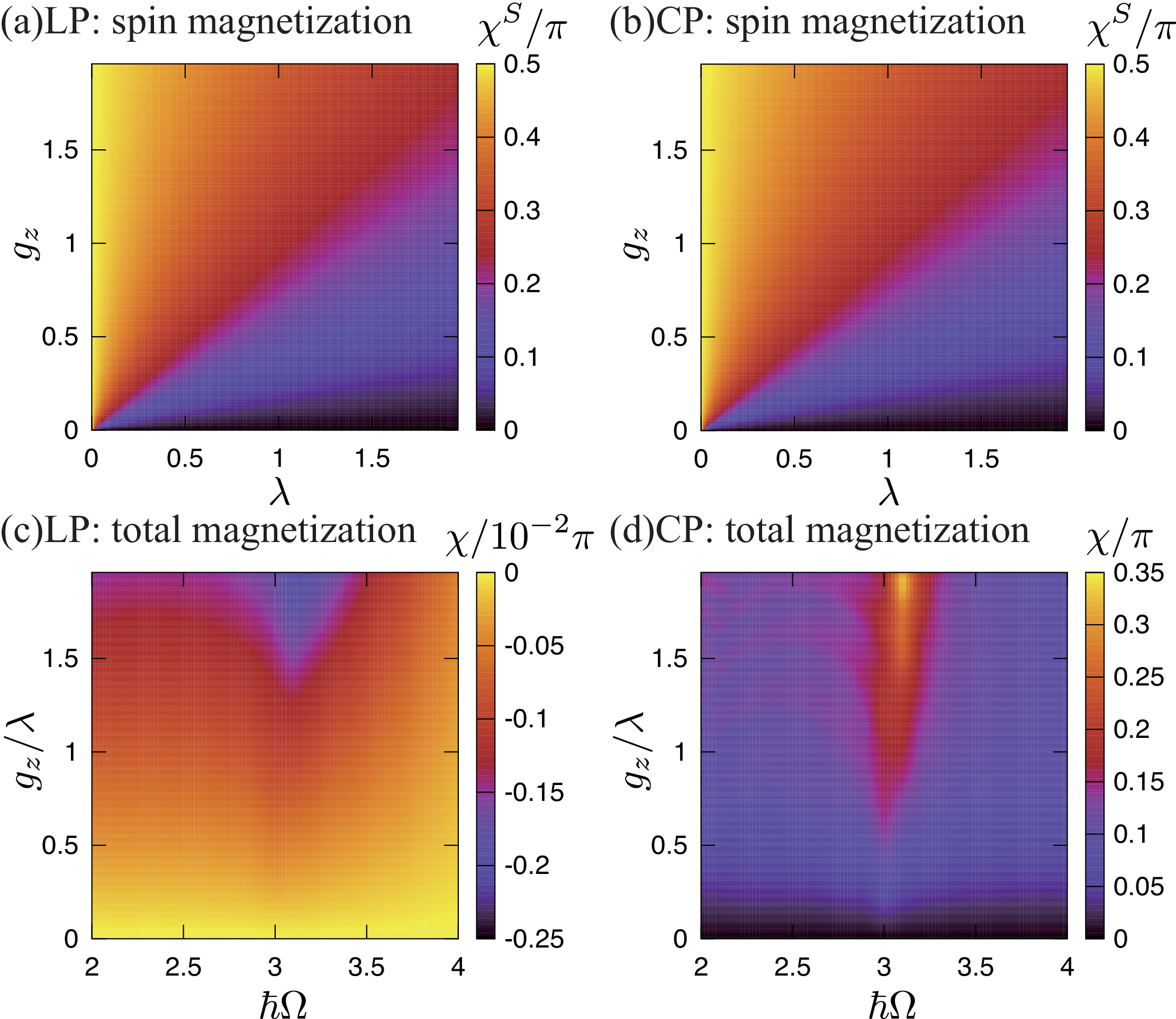}
  \caption{Angle $\chi$ in Eq.~\eqref{eq:magnetization_angle}.
  (a) and (b) SOC ($\lambda$ and $g_z$) dependence for the spin magnetization at $\hbar\Omega=3$.
  (c) and (d) Dependence on frequency and the SOC ratio for the total magnetization at $\lambda=0.4$.}
  \label{fig:magnetization_angle}
\end{figure*}

In the ferroaxial state, the induced magnetization is tilted from the light direction owing to $\delta M_{\perp}$.
To discuss its direction, we evaluate the angle
\begin{equation}
  \tan{\chi}\equiv\frac{\delta M_{\perp}}{\delta M_{\parallel}}
  =
  \begin{cases}
    \eta_{y;yz}/\eta_{x;yz} & (\mathrm{LP})
    \\
    \xi_{y;x}/\xi_{x;x} & (\mathrm{CP})
  \end{cases},
  \label{eq:magnetization_angle}
\end{equation}
with $-\pi/2<\chi\leq\pi/2$.
$\chi=0$ $(\pi/2)$ represents magnetization along the $x$ ($y$) axis.
We begin by focusing solely on the spin magnetization.
Figures~\ref{fig:magnetization_angle}(a) and~\ref{fig:magnetization_angle}(b) show two types of the SOC dependence of the angle of the spin magnetization $\delta M^{S}$ induced by the LPL and CPL at $\hbar\Omega=3$, respectively.
In the normal state where $g_{z}=0$, $\delta M^{S}$ is along the direction of the incident light, i.e., the $x$ axis.
Upon the onset of ferroaxial ordering, $\delta M^{S}$ begins to tilt with the development of the ferroaxial-order-driven SOC.
In the limit of $g_{z}/\lambda\to \infty$, $\delta M^{S}$ is perpendicular to the direction of the light, aligning along the $y$ axis.
We confirm that the angle is independent of the frequency of the light; particularly the following relation holds:
\begin{equation}
  \tan{\chi^{S}}=\frac{\delta M_{\perp}^{S}}{\delta M_{\parallel}^{S}}=\frac{g_{z}}{\lambda},
\end{equation}
for both the LPL- and CPL-induced NLEE in all parameter ranges.
This finding implies that the ratio of the ferroaxial-origin SOC to the relativistic SOC can be experimentally extracted from the measurement of the spin magnetization such as those conducted via the magnetic Compton scattering~\cite{aguiMicroscopicMagnetizationProcess2011,itouSpinOrbitalMagnetization2013}.

Figures~\ref{fig:magnetization_angle}(c) and~\ref{fig:magnetization_angle}(d) show the angle of the total magnetization $\delta M=\delta M^{L}+\delta M^{S}$, where we set $\lambda=0.4$.
Although our formalism does not allow for quantitative estimation of the orbital magnetization itself~\cite{restaOrbitalMagnetizationExtended2005,thonhauserOrbitalMagnetizationPeriodic2005,xiaoBerryPhaseCorrection2005,ceresoliOrbitalMagnetizationCrystalline2006,shiQuantumTheoryOrbital2007,souzaDichroic$f$sumRule2008,xiaoBerryPhaseEffects2010,thonhauserTheoryOrbitalMagnetization2011}, the calculated results still reveal that the total magnetization is tilted away from the direction of the incident light due to the ferroaxial order. 
This angular deviation implies that the presence of the ferroaxial order can be experimentally identified by measuring the direction of the total magnetization.
Since the NLEE does not require SI-symmetry breaking, the induced magnetization contains no contribution from the first-order response to the electric field, i.e., the LEE.
As a result, the angle defined by Eq.~\eqref{eq:magnetization_angle} is independent of the magnitude of the external field, unlike the third-order nonlinear transverse magnetization~\cite{indaNonlinearTransverseMagnetic2023}.

\section{Summary and application to other MPGs}
\label{sec:summary}

We investigated the light-induced NLEE in the presence of ferroaxial ordering.
Symmetry analysis revealed that both LPL- and CPL-induced NLEE can arise under ferroaxial order, although their transition processes differ between the two cases.
For practical use, Table~\ref{tab:process_NLEE} summarizes which polarization channel probes which microscopic processes.
The LPL-induced tensor $\eta_{\mu;\nu\lambda}$ couples to the symmetric bilinear
$L_{\nu\lambda}(\Omega)=\mathrm{Re}[E_\nu(\Omega)E_\lambda^*(\Omega)]$ and, in TR-symmetric insulators, is typically dominated by resonant (absorptive) interband contributions.
In contrast, the CPL-induced tensor $\xi_{\mu;\rho}$ couples to the axial bilinear
$F_\rho(\Omega)=-(i/2)[\mathbf{E}(\Omega)\times\mathbf{E}^*(\Omega)]_\rho$ and includes an injection-type contribution proportional to the relaxation time.
Based on a minimal model, we found that the $xy$ component of the spin magnetization is generated via the cross-product SOC characteristic of ferroaxial order.
Measuring the magnetization orientation thus enables detection of ferroaxial ordering.
Especially, the ratio of ferroaxial-origin SOC to relativistic SOC can be quantitatively extracted from the spin-magnetization angle.

Our result can apply to a broader class of systems characterized by different crystallographic point groups.
Since the NLEE occurs irrespective of the presence or absence of SI and TR symmetry, ferroaxial-origin NLEE can be found even in systems where SI or TR symmetry is already broken.
Although LEE contributions exist when SI symmetry is broken, their contributions can be separated by the difference in frequency dependence as far as the light-induced response is concerned;
the electric field $\mathbf{E}(\omega)$ induces $\delta M(\omega)$ via the LEE, $\delta M(2\omega)$ via the SHG, and $\delta M(0)$ via the NLEE.
Moreover, the NLEE can be detected in antiferromagnetic materials where ferroaxial and magnetic order parameters coexist.
Using the polarization/transition-process guide in Table~\ref{tab:process_NLEE}, we now summarize the symmetry-allowed NLEE components for MPGs.
Table~\ref{tab:mpg_nlee} lists, for MPGs without a magnetic dipole, which components of $\eta$ (LPL) and/or $\xi$ (CPL) can be nonzero, so that the ferroaxial-derived NLEE can be isolated from other multipole-driven contributions.
In particular, magnetic point groups hosting a ferroaxial vector allow characteristic transverse components (e.g., $\xi_{x;y}=-\xi_{y;x}$ and $\eta_{z;xx}=\eta_{x;zx}$), whose sign reversal provides a direct indicator of opposite ferroaxial domains.

\begin{table}[t]
  \centering
  \caption{Magnetic point groups (MPGs) that host the ferroaxial-derived NLEE but without the magnetic dipole in the presence $(\bigcirc)$ or absence $(\times)$ of the spatial inversion $\mathcal{P}$, the time-reversal $\mathcal{T}$, and their product $\mathcal{PT}$ symmetry.
  The second column shows the presence $(\checkmark)$ of the linear Edelstein effect (LEE).
  The third column represents the component of the electric toroidal (ET) dipole.}
  \begin{tabular}{ccll}
    \hline\hline
    $(\mathcal{P},\mathcal{T},\mathcal{PT})$ & LEE & ET dipole & MPGs 
    \\
    \hline
    $(\bigcirc,\bigcirc,\bigcirc)$ & --- & $G_{z}$ & $4/m1^{\prime}$, $6/m1^{\prime}$, $\bar{3}1^{\prime}$
    \\
    & & $G_{y}$ & $2/m1^{\prime}$
    \\
    & & $G_{x},G_{y},G_{z}$ & $\bar{1}1^{\prime}$
    \\
    \hline
    $(\bigcirc,\times,\times)$ & --- & $G_{z}$ & $4^{\prime}/m$, $6^{\prime}/m^{\prime}$
    \\
    \hline
    $(\times,\bigcirc,\times)$ & $\checkmark$ & $G_{z}$ & $41^{\prime}$, $\bar{4}1^{\prime}$, $61^{\prime}$, $\bar{6}1^{\prime}$, $31^{\prime}$
    \\
    & & $G_{y}$ & $21^{\prime}$, $m1^{\prime}$
    \\
    & & $G_{x},G_{y},G_{z}$ & $11^{\prime}$
    \\
    \hline
    $(\times,\times,\bigcirc)$ & $\checkmark$ & $G_{z}$ & $4^{\prime}/m^{\prime}$, $4/m^{\prime}$, $6^{\prime}/m$, $6/m^{\prime}$, $\bar{3}^{\prime}$
    \\
    & & $G_{y}$ & $2^{\prime}/m$, $2/m^{\prime}$
    \\
    & & $G_{x},G_{y},G_{z}$ & $\bar{1}^{\prime}$
    \\
    \hline
    $(\times,\times,\times)$ & $\checkmark$ & $G_{z}$ & $4^{\prime}$, $\bar{4}^{\prime}$, $6^{\prime}$, $\bar{6}^{\prime}$
    \\
    \hline\hline
  \end{tabular}
  \label{tab:mpg_nlee}
\end{table}

\acknowledgements{
  This research was supported by JSPS KAKENHI Grants No. JP21H01037, No. JP22H00101, No. JP22H01183, No. JP23K03288, No. JP23H04869, and No. JP23K20827, by JST CREST (No. JPMJCR23O4), and by JST FOREST (No. JPMJFR2366).
  The numerical calculations were performed in the supercomputing systems in ISSP, the University of Tokyo.
}

\appendix

\section{Derivation of the NLEE tensor}
\label{sec:derivation}

In this appendix, we present the details of the derivation of the NLEE tensor.
The derivation is based on the time-evolution of the reduced density matrix~\cite{watanabeChiralPhotocurrentParityViolating2021,xuLightinducedStaticMagnetization2021,oikeNonlinearMagnetoelectricEffect2024}.

The non-interacting Hamiltonian is given by 
\begin{equation}
  H_{0}=\sum_{\mathbf{k}a}\varepsilon_{\mathbf{k}a}c_{\mathbf{k}a}^{\dag}c_{\mathbf{k}a},
\end{equation}
where we have defined the creation and annihilation operators $c_{\mathbf{k}a}^{\dag},c_{\mathbf{k}a}$ of the Bloch state labeled by the crystal momentum $\mathbf{k}$ and band index $a$.
We denote the periodic part of the Bloch state for the band $\varepsilon_{\mathbf{k}a}$ by $\ket{\mathbf{k}a}$.

Next, we consider the interaction between electrons and electromagnetic fields.
We apply the dipole approximation, where the electromagnetic field is described by a uniform electric field $\mathbf{E}(t)$, and choose the length-gauge approach.
The velocity gauge approach yields the same result as the length gauge, just as in the case of the photocurrent~\cite{holderConsequencesTimereversalsymmetryBreaking2020,kaplanUnifyingSemiclassicsQuantum2023,parkerDiagrammaticApproachNonlinear2019}.
Thus, the external field is given by 
\begin{equation}
  V(t)=e\sum_{\mathbf{k}ab}\mathbf{r}_{\mathbf{k}ab}\cdot\mathbf{E}(t)c_{\mathbf{k}a}^{\dag}c_{\mathbf{k}b}. 
\end{equation}
Here, $e(>0)$ is the elementary charge.
The Bloch representation of the position operator $\mathbf{r}_{\mathbf{k}ab}$ consists of the derivative of crystal momentum $\nabla_{\mathbf{k}}$ and the Berry connection $\mathbf{\xi}_{\mathbf{k}ab}=i\braket{\mathbf{k}a|\nabla_{\mathbf{k}}|\mathbf{k}b}$, that is, 
\begin{equation}
  \mathbf{r}_{\mathbf{k}ab}=i\nabla_{\mathbf{k}}\delta_{ab}+\mathbf{\xi}_{\mathbf{k}ab}.
  \label{eq:position_op_Bloch_state}
\end{equation}
We divide the Berry connection as 
\begin{equation}
  \mathbf{\xi}_{\mathbf{k}ab}=\mathbf{\alpha}_{\mathbf{k}ab}+\mathbf{A}_{\mathbf{k}ab},
  \label{eq:berry_connection}
\end{equation}
where the intraband Berry connection $\mathbf{\alpha}_{\mathbf{k}ab}$ is introduced for the degenerate bands satisfying $\varepsilon_{\mathbf{k}a}=\varepsilon_{\mathbf{k}b}$, whereas $\mathbf{A}_{\mathbf{k}ab}$ is the interband Berry connection between non-degenerate states $\varepsilon_{\mathbf{k}a}\neq\varepsilon_{\mathbf{k}b}$.
For the latter case, the position operator can be expressed as 
\begin{equation}
  \mathbf{r}_{\mathbf{k}ab}=\mathbf{A}_{\mathbf{k}ab}=-i\hbar\frac{\mathbf{v}_{\mathbf{k}ab}}{\varepsilon_{\mathbf{k}ab}},
  \label{eq:inter_berry_connection}
\end{equation}
with $\varepsilon_{\mathbf{k}ab}=\varepsilon_{\mathbf{k}a}-\varepsilon_{\mathbf{k}b}$.
Here, $\hbar$ is Dirac constant and $\mathbf{v}_{\mathbf{k}ab}$ is the Bloch representation of the velocity operator $\hat{\mathbf{v}}_{\mathbf{k}}=\hbar^{-1}\nabla_{\mathbf{k}}\hat{h}(\mathbf{k})$.
The gauge covariant derivative is defined by 
\begin{equation}
  [\mathbf{D}O]_{ab}=\nabla_{\mathbf{k}}O_{ab}-i[\mathbf{\alpha}_{\mathbf{k}},O]_{ab},
\end{equation}
where $[A,B]_{ab}=\sum_{c}[A_{ac}B_{cb}-B_{ac}A_{cb}]$.

For convenience, we also introduce the $h$-space Berry connection~\cite{oikeNonlinearMagnetoelectricEffect2024}
\begin{equation}
  \mathbf{\xi}^{(h)}_{\mathbf{k}ab}=i\braket{\mathbf{k}a;\mathbf{h}|\nabla_{\mathbf{h}}|\mathbf{k}b;\mathbf{h}}_{\mathbf{h}=\mathbf{0}},
  \label{eq:h_space_berry_connection}
\end{equation}
where $\ket{\mathbf{k}a;\mathbf{h}}$ is the Bloch state for the Hamiltonian 
\begin{equation*}
  H(\mathbf{h})=H_{0}-\sum_{\mathbf{k}ab}\mathbf{\beta}_{ab}\cdot\mathbf{h}c_{\mathbf{k}a}^{\dag}c_{\mathbf{k}b},
\end{equation*}
with the fictional external field $\mathbf{h}$.
Here, we assume that the NLEE tensor for orbital magnetization can be derived in the same way as for spin magnetization.
As with the Berry connection in Eq.~\eqref{eq:berry_connection}, we can decompose $\mathbf{\xi}^{(h)}_{\mathbf{k}ab}$ into the intraband one $\mathbf{\alpha}^{(h)}_{\mathbf{k}ab}$ for $\varepsilon_{\mathbf{k}a}=\varepsilon_{\mathbf{k}b}$ and the interband one $\mathbf{A}^{(h)}_{\mathbf{k}ab}$ for $\varepsilon_{\mathbf{k}a}\neq\varepsilon_{\mathbf{k}b}$.
In particular, the interband $h$-space Berry connection can be expressed as 
\begin{equation}
  \mathbf{A}^{(h)}_{\mathbf{k}ab}=i\frac{\mathbf{\beta}_{ab}}{\varepsilon_{\mathbf{k}ab}}.
  \label{eq:h_space_inter_berry_connection}
\end{equation}
$\mathbf{\beta}_{ab}$ is the Bloch representation of the angular momentum operator.
The $h$-space covariant derivative is defined by 
\begin{equation}
  [\mathbf{D}^{(h)}O]_{ab}=\nabla_{\mathbf{h}}O_{ab}-i[\mathbf{\alpha}^{(h)}_{\mathbf{k}},O]_{ab}.
\end{equation}
From the definitions of $\mathbf{D}$ and $\mathbf{D}^{(h)}$, we can derive the sum rule~\cite{watanabeChiralPhotocurrentParityViolating2021,oikeNonlinearMagnetoelectricEffect2024}
\begin{equation}
  [D^{(h)\mu}A^{\nu}_{\mathbf{k}}]_{ab}-[D^{\nu}A^{(h)\mu}_{\mathbf{k}}]_{ab}=i[A^{(h)\mu}_{\mathbf{k}},A^{\nu}_{\mathbf{k}}]_{ab}.
  \label{eq:sum_rule}
\end{equation}

\subsection{Dynamics of the density operator}

The dynamics of the density operator $\rho(t)$ obeys the von Neumann equation:
\begin{equation}
  i\hbar\partial_{t}\rho(t)=[H(t),\rho(t)],
  \label{eq:von_Neumann_eq_1}
\end{equation}
where $\partial_{t}=\partial/\partial t$, $[A,B]=AB-BA$, and $H(t)=H_{0}+V(t)$.
The density operator $\rho(t)$ is expressed by the tensor product labeled by the crystal momentum $\mathbf{k}$, that is, $\rho(t)=\prod_{\mathbf{k}}\otimes\rho_{\mathbf{k}}(t)$, and the matrix representation of $\rho_{\mathbf{k}}(t)$ is defined as 
\begin{equation}
  \rho_{\mathbf{k}ab}(t)=\mathrm{Tr}[\rho(t)c_{\mathbf{k}b}^{\dag}c_{\mathbf{k}a}].
\end{equation}
Thus, the von Neumann equation in Eq.~\eqref{eq:von_Neumann_eq_1} is rewritten as 
\begin{equation}
  (i\hbar\partial_{t}-\varepsilon_{\mathbf{k}ab})\rho_{\mathbf{k}ab}(t)=e\mathbf{E}(t)\cdot[\mathbf{r}_{\mathbf{k}},\rho_{\mathbf{k}}(t)]_{ab}.
  \label{eq:von_Neumann_eq_2}
\end{equation}
Adopting the Fourier transformation, we obtain Eq.~\eqref{eq:von_Neumann_eq_2} in frequency domain as
\begin{equation}
  (\hbar\omega-\varepsilon_{\mathbf{k}ab})\rho_{\mathbf{k}ab}(\omega)=e\int_{-\infty}^{\infty}\frac{d\Omega}{2\pi}\mathbf{E}(\Omega)\cdot[\mathbf{r}_{\mathbf{k}},\rho_{\mathbf{k}}(\omega-\Omega)]_{ab}.
  \label{eq:von_Neumann_eq_w}
\end{equation}

Now we expand $\rho_{\mathbf{k}}(t)$ in powers of the magnitude of electric field $|\mathbf{E}|$ such as $\rho_{\mathbf{k}}(t)=\sum_{l}\rho_{\mathbf{k}}^{(l)}(t)$ with $\rho_{\mathbf{k}}^{(l)}=O(|\mathbf{E}|^{l})$.
Furthermore, we phenomenologically introduce a scattering term into Eq.~\eqref{eq:von_Neumann_eq_w}.
Since the effective relaxation time decreases with higher orders in perturbation theory, we assume that the scattering rate from $O(|\mathbf{E}|^{l})$ is proportional to $l$
~\cite{chengThirdorderNonlinearityGraphene2015,PhysRevB.97.235446,watanabeChiralPhotocurrentParityViolating2021,dasIntrinsicNonlinearConductivities2023,oikeNonlinearMagnetoelectricEffect2024}.
Therefore, we obtain the recursive equation
\begin{equation}
  (\hbar\omega-\varepsilon_{\mathbf{k}ab}+il\gamma)\rho_{\mathbf{k}ab}^{(l)}(\omega)=e\int_{-\infty}^{\infty}\frac{d\Omega}{2\pi}\mathbf{E}(\Omega)\cdot[\mathbf{r}_{\mathbf{k}},\rho_{\mathbf{k}}^{(l-1)}(\omega-\Omega)]_{ab}
  \label{eq:von_Neumann_eq_lth}
\end{equation}
for $l=1,2,\ldots$, where $\gamma$ is the scattering rate.
The zeroth component is given by $\rho_{\mathbf{k}ab}^{(0)}(\omega)=2\pi\delta(\omega)\delta_{ab}f_{\mathbf{k}a}$ with the Fermi distribution function $f_{\mathbf{k}a}$ for the Bloch state $\ket{\mathbf{k}a}$.

Let us solve Eq.~\eqref{eq:von_Neumann_eq_lth}.
For $l=1$, the solution is 
\begin{equation}
  \rho^{(1)}_{\mathbf{k}ab}(\omega)
  =\left[ie\delta_{ab}\frac{\nabla_{\mathbf{k}}f_{\mathbf{k}a}}{\hbar\omega+i\gamma}
  -e\frac{\mathbf{A}_{\mathbf{k}ab}f_{\mathbf{k}ab}}{\hbar\omega-\varepsilon_{\mathbf{k}ab}+i\gamma}\right]\cdot\mathbf{E}(\omega),
\end{equation}
where we have used Eqs.~\eqref{eq:position_op_Bloch_state} and \eqref{eq:berry_connection} and $f_{\mathbf{k}ab}=f_{\mathbf{k}a}-f_{\mathbf{k}b}$.
Note that the contribution from the intraband Berry connection $\mathbf{\alpha}_{ab}$ vanishes by $f_{\mathbf{k}ab}=0$.
Thus, the first-order density matrix is decomposed into the intraband (J) and interband (E) contributions, i.e., 
\begin{subequations}
  \begin{equation}
    \rho_{\mathbf{k}ab}^{(1\mathrm{J})}(\omega)=ie\delta_{ab}d^{(1)}_{\omega,\mathbf{k}aa}\mathbf{E}(\omega)\cdot\nabla_{\mathbf{k}}f_{\mathbf{k}a},
  \end{equation}
  \begin{equation}
    \rho_{\mathbf{k}ab}^{(1\mathrm{E})}(\omega)=-ed^{(1)}_{\omega,\mathbf{k}ab}\mathbf{E}(\omega)\cdot\mathbf{A}_{\mathbf{k}ab}f_{\mathbf{k}ab}.
  \end{equation}
  \label{eq:1st_density_matrix}
\end{subequations}
Here, we have introduced 
\begin{equation}
  d^{(l)}_{\omega,\mathbf{k}ab}=\frac{1}{\hbar\omega-\varepsilon_{\mathbf{k}ab}+il\gamma}.
\end{equation}

The second-order density matrix is further decomposed into intraband and interband contributions as 
\begin{equation*}
  \begin{aligned}
    \rho^{(2\mathrm{J})}_{\mathbf{k}ab}(\omega)=&\, d^{(2)}_{\omega,\mathbf{k}ab}e\int_{-\infty}^{\infty}\frac{d\Omega}{2\pi}\mathbf{E}(\Omega)\cdot[\mathbf{r}_{\mathbf{k}},\rho^{(1\mathrm{J})}_{\mathbf{k}}(\omega-\Omega)]_{ab}
    \\
    =&\, \rho^{(2\mathrm{JJ})}_{\mathbf{k}ab}(\omega)+\rho^{(2\mathrm{EJ})}_{\mathbf{k}ab}(\omega),
    \\
    \rho^{(2\mathrm{E})}_{\mathbf{k}ab}(\omega)=&\, d^{(2)}_{\omega,\mathbf{k}ab}e\int_{-\infty}^{\infty}\frac{d\Omega}{2\pi}\mathbf{E}(\Omega)\cdot[\mathbf{r}_{\mathbf{k}},\rho^{(1\mathrm{E})}_{\mathbf{k}}(\omega-\Omega)]_{ab}
    \\
    =&\, \rho^{(2\mathrm{JE})}_{\mathbf{k}ab}(\omega)+\rho^{(2\mathrm{EE})}_{\mathbf{k}ab}(\omega),
  \end{aligned}
\end{equation*}
where
\begin{widetext}
  \begin{subequations}
    \begin{equation}
      \rho^{(2\mathrm{JJ})}_{\mathbf{k}ab}(\omega)=-e^{2}\delta_{ab}\int_{-\infty}^{\infty}\frac{d\Omega}{2\pi}E_{\mu}(\Omega)E_{\nu}(\omega-\Omega)d_{\omega,\mathbf{k}aa}^{(2)}d_{\omega-\Omega,\mathbf{k}aa}^{(1)}\partial^{\mu}\partial^{\nu}f_{\mathbf{k}a},
      \label{eq:2nd_density_matrix_jj}
    \end{equation}
    \begin{equation}
      \rho^{(2\mathrm{EJ})}_{\mathbf{k}ab}(\omega)=-ie^{2}\int_{-\infty}^{\infty}\frac{d\Omega}{2\pi}E_{\mu}(\Omega)E_{\nu}(\omega-\Omega)d_{\omega,\mathbf{k}ab}^{(2)}d_{\omega-\Omega,\mathbf{k}aa}^{(1)}A^{\mu}_{\mathbf{k}ab}\partial^{\nu}f_{\mathbf{k}ab},
      \label{eq:2nd_density_matrix_ej}
    \end{equation}
    \begin{equation}
      \rho^{(2\mathrm{JE})}_{\mathbf{k}ab}(\omega)
      =-e^{2}\int_{-\infty}^{\infty}\frac{d\Omega}{2\pi}E_{\mu}(\Omega)E_{\nu}(\omega-\Omega)d^{(2)}_{\omega,\mathbf{k}ab}
      \left\{i\partial^{\mu}(d^{(1)}_{\omega-\Omega,\mathbf{k}ab}A^{\nu}_{\mathbf{k}ab}f_{\mathbf{k}ab})+d^{(1)}_{\omega-\Omega,\mathbf{k}ab}f_{\mathbf{k}ab}[\alpha^{\mu}_{\mathbf{k}},A^{\nu}_{\mathbf{k}}]_{ab}\right\},
      \label{eq:2nd_density_matrix_je}
    \end{equation}
    \begin{equation}
      \rho^{(2\mathrm{EE})}_{\mathbf{k}ab}(\omega)
      =-e^{2}\int_{-\infty}^{\infty}\frac{d\Omega}{2\pi}E_{\mu}(\Omega)E_{\nu}(\omega-\Omega)d^{(2)}_{\omega,\mathbf{k}ab}
      \sum_{c}\left[d^{(1)}_{\omega-\Omega,\mathbf{k}cb}A^{\mu}_{\mathbf{k}ac}A^{\nu}_{\mathbf{k}cb}f_{\mathbf{k}cb}-d^{(1)}_{\omega-\Omega,\mathbf{k}ac}A^{\mu}_{\mathbf{k}cb}A^{\nu}_{\mathbf{k}ac}f_{\mathbf{k}ac}\right],
      \label{eq:2nd_density_matrix_ee}
    \end{equation}
    \label{eq:2nd_density_matrix}
  \end{subequations}
\end{widetext}
with $\partial^{\mu}=\partial/\partial k_{\mu}$.
We have omitted the summation over $\mu$ and $\nu$ in Eq.~\eqref{eq:2nd_density_matrix}.
Finally, we symmetrize the indices and frequencies of the electric fields.

\subsection{Derivation of NLEE tensor}

The second-order induced magnetization per unit cell is defined by 
\begin{equation}
  \delta M_{\mu}^{\beta}(\omega)=\mu_{\mathrm{B}}\frac{1}{N}\sum_{\mathbf{k}ab}\beta^{\mu}_{\mathbf{k}ba}\rho^{(2)}_{\mathbf{k}ab}(\omega),
\end{equation}
for $\mu=x,y,z$.
Here, $\mu_{\mathrm{B}}$ is the Bohr magneton, and $\beta$ represents the spin $\beta=S$ or orbital $\beta=L$.
Using Eq.~\eqref{eq:2nd_density_matrix} with symmetrization for electric fields, we obtain the NLEE tensor 
\begin{equation}
  \delta M_{\mu}(\omega)=\sum_{\nu\lambda}\int\frac{d\omega_{1}d\omega_{2}}{(2\pi)^{2}}\tilde{\zeta}_{\mu;\nu\lambda}(\omega;\omega_{1},\omega_{2})E_{\nu}(\omega_{1})E_{\lambda}(\omega_{2}),
  \label{eq:2nd_induced_M}
\end{equation}
where 
\begin{equation}
  \tilde{\zeta}_{\mu;\nu\lambda}(\omega;\omega_{1},\omega_{2})=2\pi\delta(\omega-\omega_{1}-\omega_{2})\zeta_{\mu;\nu\lambda}(\omega_{1},\omega_{2}).
\end{equation}
We have omitted the superscript $\beta$.
In the following, we consider the static magnetization induced by the dynamical electric fields, that is, $\omega=0,\omega_{1}=-\omega_{2}=\pm\Omega$ with $\Omega>0$.
We also consider the optical regime $\hbar\Omega\gg \gamma$ and that the band gap is larger than the scattering rate $\varepsilon_{ab}\gg\gamma$.

\subsubsection{JJ term}

Hereafter, we shall omit the $\mathbf{k}$ index and $\frac{1}{N}\sum_{\mathbf{k}}$ from the expressions for simplicity.
The JJ term coming from Eq.~\eqref{eq:2nd_density_matrix_jj} is 
\begin{equation}
  \zeta_{\mu;\nu\lambda}^{(\mathrm{JJ})}(\Omega)
  =\frac{\mu_{\mathrm{B}}e^{2}}{2(\hbar\Omega)^{2}}\sum_{a}\beta^{\mu}_{aa}\partial^{\nu}\partial^{\lambda}f_{a}.
\end{equation}
Here, $\zeta_{\mu;\nu\lambda}(\Omega)=\zeta_{\mu;\nu\lambda}(\Omega,-\Omega)$.
Clearly, the JJ term is classified as the LPL-induced NLEE in metals, that is, 
\begin{equation}
  \eta_{\mu;\nu\lambda}^{(\mathrm{JJ})}(\Omega)=\frac{\mu_{\mathrm{B}}e^{2}}{2(\hbar\Omega)^{2}}\sum_{a}\beta^{\mu}_{aa}\partial^{\nu}\partial^{\lambda}f_{a}.
  \label{eq:nlee_jj}
\end{equation}

Let us apply the TR operation $\mathcal{T}$ to Eq.~\eqref{eq:nlee_jj}.
Since $\mathcal{T}:\beta_{aa}^{\mu}\mapsto -\beta_{aa}^{\mu}$, the JJ term has only the TR-odd part related to the magnetic/MT multipoles.

\subsubsection{EJ term}

Second, the EJ term derived from Eq.~\eqref{eq:2nd_density_matrix_ej} is 
\begin{equation}
  \begin{aligned}
    \zeta_{\mu;\nu\lambda}^{(\mathrm{EJ})}(\Omega)
    =&\, \frac{\mu_{\mathrm{B}}e^{2}}{2i}\sum_{ab}\beta^{\mu}_{ba}d_{0,ab}^{(2)}d_{-\Omega,aa}^{(1)}A^{\nu}_{ab}\partial^{\lambda}f_{ab}
    +(\nu,\Omega)\leftrightarrow(\lambda,-\Omega)
    \\
    =&\, -i\frac{\mu_{\mathrm{B}}e^{2}}{2\hbar\Omega}\sum_{a}\partial^{\lambda}\Upsilon_{a}^{\mu\nu}f_{a}+(\nu,\Omega)\leftrightarrow(\lambda,-\Omega).
  \end{aligned}
\end{equation}
We have introduced the anomalous spin/orbital polarizability~\cite{xiaoTimeReversalEvenNonlinearCurrent2023}
\begin{equation}
  \Upsilon_{a}^{\mu\nu}=-2\sum_{b}\mathrm{Im}[A_{ab}^{(h)\mu}A_{ba}^{\nu}].
\end{equation}
The EJ term is induced by the CPL in metals, that is,
\begin{equation}
  \xi_{\mu;\rho}^{(\mathrm{EJ})}(\Omega)=\frac{\mu_{\mathrm{B}}e^{2}}{\hbar\Omega}\sum_{a}\sum_{\nu\lambda}\epsilon_{\rho\nu\lambda}\partial^{\lambda}\Upsilon_{a}^{\mu\nu}f_{a}.
\end{equation}
Since $\mathcal{T}:\Upsilon_{a}^{\mu\nu}\mapsto -\Upsilon_{a}^{\mu\nu}$, the EJ term has only the TR-even part related to the electric/ET multipoles.
This contribution is analogous to the Berry curvature dipole term in the photocurrent~\cite{watanabeChiralPhotocurrentParityViolating2021}.

\subsubsection{Intraband EE term}

We next consider the EE term, particularly, the intraband component of the angular momentum operator $\beta_{ab}^{\mu}$ ($\varepsilon_{a}=\varepsilon_{b}$) with Eq.~\eqref{eq:2nd_density_matrix_ee}.
The corresponding contribution $\zeta_{\mu;\nu\lambda}^{(\mathrm{EE;d})}$ is given by 
\begin{equation}
  \begin{aligned}
    \zeta_{\mu;\nu\lambda}^{(\mathrm{EE;d})}(\Omega)
    =&\, \frac{\mu_{\mathrm{B}}e^{2}}{2}\sum_{abc}^{\varepsilon_{c}=\varepsilon_{a}}\beta^{\mu}_{ca}d^{(2)}_{0,aa}
    \left[d^{(1)}_{-\Omega,ba}A^{\nu}_{ab}A^{\lambda}_{bc}+d^{(1)}_{-\Omega,ab}A^{\lambda}_{ab}A^{\nu}_{bc}\right]f_{ab}
    \\
    &\, +(\nu,\Omega)\leftrightarrow(\lambda,-\Omega)
    \\
    =&\, -\frac{\mu_{\mathrm{B}}e^{2}}{2\gamma}\sum_{ab}
    \frac{\gamma}{(\hbar\Omega-\varepsilon_{ab})^{2}+\gamma^{2}}[\beta^{\mu},A^{\nu}]_{ab}^{\prime}A^{\lambda}_{ba}f_{ab}.
  \end{aligned}
\end{equation}
Here, we have introduced
\begin{equation*}
  [\beta^{\mu},A^{\nu}]_{ab}^{\prime}\equiv\sum_{c}^{\varepsilon_{c}=\varepsilon_{a}}\beta_{ac}^{\mu}A^{\nu}_{cb}-\sum_{c}^{\varepsilon_{c}=\varepsilon_{b}}A^{\nu}_{ac}\beta_{cb}^{\mu}.
\end{equation*}
This term is proportional to the relaxation time $\tau=\hbar/\gamma$ under the clean limit $\gamma\to0$:
\begin{equation}
  \zeta_{\mu;\nu\lambda}^{(\mathrm{EE;d})}(\Omega)\to -\frac{\pi\mu_{\mathrm{B}}e^{2}\tau}{2\hbar}\sum_{ab}\delta(\hbar\Omega-\varepsilon_{ab})[\beta^{\mu},A^{\nu}]_{ab}^{\prime}A^{\lambda}_{ba}f_{ab}.
\end{equation}
This term has both LPL- and CPL-induced NLEE components:
\begin{subequations}
  \label{eq:nlee_eed}
  \begin{equation}
    \eta^{(\mathrm{EE;d})}_{\mu;\nu\lambda}(\Omega)=-\frac{\pi\mu_{\mathrm{B}}e^{2}\tau}{2\hbar}\sum_{ab}\delta(\hbar\Omega-\varepsilon_{ab})\mathrm{Re}\left([\beta^{\mu},A^{\nu}]_{ab}^{\prime}A^{\lambda}_{ba}\right)f_{ab},
    \label{eq:magnetic_spin_injection}
  \end{equation}
  \begin{equation}
    \xi^{(\mathrm{EE;d})}_{\mu;\rho}(\Omega)=\frac{\pi\mu_{\mathrm{B}}e^{2}\tau}{2\hbar}\sum_{ab}\delta(\hbar\Omega-\varepsilon_{ab})\sum_{\nu\lambda}\epsilon_{\rho\nu\lambda}\mathrm{Im}\left([\beta^{\mu},A^{\nu}]_{ab}^{\prime}A_{ba}^{\lambda}\right)f_{ab}.
    \label{eq:electric_spin_injection}
  \end{equation}
\end{subequations}
Here, $\eta^{(\mathrm{EE;d})}$ and $\xi^{(\mathrm{EE;d})}$ are analogous to the magnetic- and electric-injection current in photocurrent~\cite{watanabeChiralPhotocurrentParityViolating2021}.
Equation~\eqref{eq:electric_spin_injection} corresponds to Eq.~\eqref{eq:cp_nlee_insulator}.
Equation~\eqref{eq:magnetic_spin_injection} is the TR-odd part, whereas Eq.~\eqref{eq:electric_spin_injection} is the TR-even part.

\subsubsection{JE and interband EE terms}

Finally, we analyze the remaining terms, that is, the contributions from Eq.~\eqref{eq:2nd_density_matrix_ej} and the interband component of the angular momentum operator $\beta_{ab}^{\mu}$ ($\varepsilon_{a}\neq\varepsilon_{b}$) with Eq.~\eqref{eq:2nd_density_matrix_ee}.
We denote the latter term by $\zeta^{(\mathrm{EE;o})}$.
The JE term is rewritten as 
\begin{equation}
  \zeta_{\mu;\nu\lambda}^{(\mathrm{JE})}(\Omega)
  =\frac{\mu_{\mathrm{B}}e^{2}}{2}\sum_{ab}d^{(1)}_{-\Omega,ba}f_{ba}\left[D^{\nu}A^{(h)\mu}\right]_{ab}A^{\lambda}_{ba}
  +(\nu,\Omega)\leftrightarrow(\lambda,-\Omega).
  \label{eq:NLEE_EJ}
\end{equation}
Meanwhile, the interband EE term is 
\begin{equation}
  \begin{aligned}
    \zeta_{\mu;\nu\lambda}^{(\mathrm{EE;o})}(\Omega)
    =&\, i\frac{\mu_{\mathrm{B}}e^{2}}{2}\sum_{ab}d^{(1)}_{-\Omega,ba}f_{ba}\left[A^{(h)\mu},A^{\nu}\right]_{ab}A^{\lambda}_{ba}
    \\
    &\, +(\nu,\Omega)\leftrightarrow(\lambda,-\Omega),
  \end{aligned}
  \label{eq:NLEE_EEo}
\end{equation}
Summing up Eqs.~\eqref{eq:NLEE_EJ} and \eqref{eq:NLEE_EEo} and using the sum rule in Eq.~\eqref{eq:sum_rule}, we obtain the simplified expression 
\begin{equation}
  \zeta_{\mu;\nu\lambda}^{(\mathrm{JE+EE;o})}(\Omega)
  =-i\frac{\mu_{\mathrm{B}}e^{2}}{2}\sum_{ab}Q_{ab}^{\mu;\nu\lambda}d^{(1)}_{-\Omega,ba}f_{ba}
  +(\nu,\Omega)\leftrightarrow(\lambda,-\Omega),
\end{equation}
where we have introduced the $h$-$k$ space geometrical connection
\begin{equation}
  Q_{ab}^{\mu;\nu\lambda}=i\left[D^{(h)\mu}A^{\nu}\right]_{ab}A^{\lambda}_{ba}.
  \label{eq:geometrical_connection}
\end{equation}
\begin{subequations}
  By taking the clean limit, the general formula for the LPL-induced NLEE tensor is
  \begin{equation}
    \begin{aligned}
      \eta_{\mu;\nu\lambda}^{(\mathrm{JE+EE;o})}(\Omega)=&\, \frac{\mu_{\mathrm{B}}e^{2}}{2}\sum_{ab}\mathrm{Im}\left(Q_{ab}^{\mu;\nu\lambda}+Q_{ab}^{\mu;\lambda\nu}\right)\frac{\varepsilon_{ab}}{(\hbar\Omega)^{2}-\varepsilon_{ab}^{2}}f_{ab}
      \\
      &\, +\frac{\pi\mu_{\mathrm{B}}e^{2}}{2}\sum_{ab}\mathrm{Re}\left(Q_{ab}^{\mu;\nu\lambda}+Q_{ab}^{\mu;\lambda\nu}\right)\delta(\hbar\Omega-\varepsilon_{ab})f_{ab},
    \end{aligned}
    \label{eq:shift_spin}
  \end{equation}
  whereas that for the CPL-induced NLEE tensor is 
  \begin{equation}
    \begin{aligned}
      \xi_{\mu;\rho}^{(\mathrm{JE+EE;o})}(\Omega)
      =&\, \mu_{\mathrm{B}}e^{2}\sum_{ab}\sum_{\nu\lambda}\epsilon_{\rho\nu\lambda}\mathrm{Re}Q_{ab}^{\mu;\nu\lambda}\frac{\hbar\Omega}{(\hbar\Omega)^{2}-\varepsilon_{ab}^{2}}f_{ab}
      \\
      &\, -\pi\mu_{\mathrm{B}}e^{2}\sum_{ab}\sum_{\nu\lambda}\epsilon_{\rho\nu\lambda}\mathrm{Im}Q_{ab}^{\mu;\nu\lambda}\delta(\hbar\Omega-\varepsilon_{ab})f_{ab}.
    \end{aligned}
    \label{eq:gyro_spin}
  \end{equation}
  The second term in Eq.~\eqref{eq:shift_spin} corresponds to Eq.~\eqref{eq:lp_nlee_insulator}.
  We have used the identity $Q_{ab}^{\mu;\nu\lambda*}=-Q_{ba}^{\mu;\nu\lambda}$.
\end{subequations} 
The absorptive part of $\eta^{(\mathrm{JE+EE;o})}_{\mu;\nu\lambda}$ [$\xi^{(\mathrm{JE+EE;o})}_{\mu;\rho}$] is analogous to the shift current (gyraion current) in photocurrent~\cite{watanabeChiralPhotocurrentParityViolating2021}.
On the other hand, the reactive  part of $\eta^{(\mathrm{JE+EE;o})}_{\mu;\nu\lambda}$ [$\xi^{(\mathrm{JE+EE;o})}_{\mu;\rho}$] corresponds to the inverse Cotton-Mouton effect~\cite{marmoElectricFieldInduced1995,kalashnikovaImpulsiveGenerationCoherent2007,kalashnikovaImpulsiveExcitationCoherent2008,ben-amarbarangaObservationInverseCottonMouton2011} (the inverse Faraday effect~\cite{vanderzielOpticallyInducedMagnetizationResulting1965,pershanTheoreticalDiscussionInverse1966, tamakiOpticallyInducedMagnetization1978,kimelUltrafastNonthermalControl2005,kirilyukUltrafastOpticalManipulation2010}).
By considering $\mathcal{T}:Q_{ab}^{\mu;\nu\lambda}\mapsto Q_{ab}^{\mu;\nu\lambda*}$, we find that the reactive (absorptive) part of Eq.~\eqref{eq:shift_spin} and the absorptive (reactive) part of Eq.~\eqref{eq:gyro_spin} are TR-odd (TR-even).
$Q_{ab}^{\mu;\nu\lambda}$ in Eq.~\eqref{eq:geometrical_connection} can be transformed as  
\begin{equation}
  Q_{ab}^{\mu;\nu\lambda}=
  i\left(\left[\beta^{\mu},A^{\nu}\right]^{\prime}_{ab}+\hbar\left[A^{(h)\mu},v^{\nu}\right]_{ab}\right)\frac{A^{\lambda}_{ba}}{\varepsilon_{ab}}.
\end{equation}
Therefore, we can evaluate it using the Bloch representation of the velocity and angular momentum operators by using the expressions for the interband Berry connection in Eqs.~\eqref{eq:inter_berry_connection} and~\eqref{eq:h_space_inter_berry_connection}.

\section{Multipole matrix elements for $p$ orbital}
\label{sec:multipoles_p_orbital}

In orbital space $(\phi_{x},\phi_{y},\phi_{z})$, any matrix in spinless space is described by the following nine spinless atomic multipoles:
\begin{equation}
  \begin{aligned}
    &\, \mathbb{Q}_{0}=\frac{1}{\sqrt{3}}\mathrm{Diag}
    \begin{pmatrix}
      1 & 1 & 1 
    \end{pmatrix},
    &\, 
    &\, \mathbb{Q}_{u}=\frac{1}{\sqrt{6}}\mathrm{Diag}
    \begin{pmatrix}
      -1 & -1 & 2 
    \end{pmatrix},
    \\
    &\, \mathbb{Q}_{v}=\frac{1}{\sqrt{2}}\mathrm{Diag}
    \begin{pmatrix}
      1 & -1 & 0 
    \end{pmatrix},
    &\, 
    &\, \mathbb{Q}_{xy}=\frac{1}{\sqrt{2}}
    \begin{pmatrix}
      0 & 1 & 0 \\
      1 & 0 & 0 \\
      0 & 0 & 0 \\
    \end{pmatrix},
    \\
    &\, \mathbb{Q}_{yz}=\frac{1}{\sqrt{2}}
    \begin{pmatrix}
      0 & 0 & 0 \\
      0 & 0 & 1 \\
      0 & 1 & 0 \\
    \end{pmatrix},
    &\, 
    &\, \mathbb{Q}_{zx}=\frac{1}{\sqrt{2}}
    \begin{pmatrix}
      0 & 0 & 1 \\
      0 & 0 & 0 \\
      1 & 0 & 0 \\
    \end{pmatrix},
    \\
    &\, \mathbb{M}_{x}=\frac{1}{\sqrt{2}}
    \begin{pmatrix}
      0 & 0 & 0 \\
      0 & 0 & -i \\
      0 & i & 0 \\
    \end{pmatrix},
    &\, 
    &\, \mathbb{M}_{y}=\frac{1}{\sqrt{2}}
    \begin{pmatrix}
      0 & 0 & i \\
      0 & 0 & 0 \\
      -i & 0 & 0 \\
    \end{pmatrix},
    \\
    &\, \mathbb{M}_{z}=\frac{1}{\sqrt{2}}
    \begin{pmatrix}
      0 & -i & 0 \\
      i & 0 & 0 \\
      0 & 0 & 0 \\
    \end{pmatrix},
  \end{aligned}
  \label{eq:mp_basis_orbital}
\end{equation}
where $\mathrm{Diag}$ means the diagonal matrix.
They are orthonormalized as $\mathrm{Tr}[\mathbb{X}_{\mu}\mathbb{X}_{\nu}]=\delta_{\mu\nu}$.
The orbital angular momentum operator $\mathbf{l}=(l_{x},l_{y},l_{z})$ is given by $l_{\mu}=\sqrt{2}\mathbb{M}_{\mu}$.
Similarly, any matrix in spinful space can be expressed as the product of the above spinless multipoles and the unit and Pauli matrices in spin space $(\uparrow,\downarrow)$ as 
\begin{equation}
  \begin{aligned}
    &\, \sigma_{0}=\mathrm{Diag}
    \begin{pmatrix}
      1 & 1 
    \end{pmatrix},
    &\, 
    &\, \sigma_{z}=\mathrm{Diag}
    \begin{pmatrix}
      1 & -1 
    \end{pmatrix},
    \\
    &\, \sigma_{x}=
    \begin{pmatrix}
      0 & 1 \\
      1 & 0 \\
    \end{pmatrix},
    &\, 
    &\, \sigma_{y}=
    \begin{pmatrix}
      0 & -i \\
      i & 0 \\
    \end{pmatrix}.
  \end{aligned}
  \label{eq:mp_basis_spin}
\end{equation}
The spin angular momentum operator $\mathbf{s}=(s_{x},s_{y},s_{z})$ is expressed as $s_{\mu}=\sigma_{\mu}/2$.

\section{Relaxation time dependence and contribution from the reactive part}
\label{sec:relaxation_time}

\begin{figure*}[htbp]
  \centering
  \includegraphics[width=\linewidth]{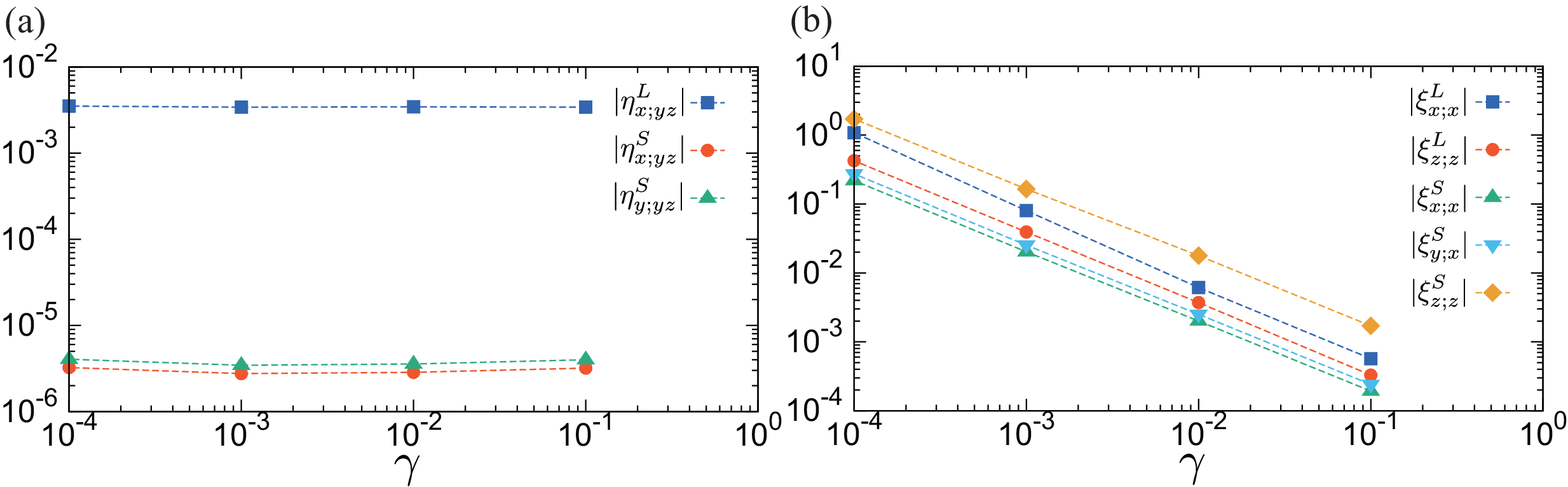}
  \caption{Scattering rate dependence of the NLEE tensor in Eq.~\eqref{eq:nlee_insulator} at $\hbar\Omega=4$.
  For the other model parameters, we use the same values as in Fig.~\ref{fig:nlee_frequency}.}
  \label{fig:relaxation_time}
\end{figure*}

Here, we see the justification of using Eq.~\eqref{eq:nlee_insulator} in the optical regime $\Omega\tau\gg1$.
Figures~\ref{fig:relaxation_time}(a) and~\ref{fig:relaxation_time}(b) show the scattering rate $\gamma$ dependence for the LPL-induced NLEE tensor in Eq.~\eqref{eq:lp_nlee_insulator} and the CPL-induced NLEE tensor in Eq.~\eqref{eq:cp_nlee_insulator} at $\hbar\Omega=4$, respectively.
We numerically confirmed that $\eta_{\mu;\nu\lambda}(\Omega)$ is almost independent of $\gamma$, whereas the absorptive part of $\xi_{\mu;\rho}(\Omega)$ linearly decreases with $\gamma^{-1}$.
Since the relaxation time is given by $\tau=\hbar/\gamma$, these results indicate that $\eta_{\mu;\nu\lambda}(\Omega)$ is independent of $\tau$, whereas $\xi_{\mu;\rho}(\Omega)$ is proportional to $\tau$ in the optical regime.

\begin{figure*}[htbp]
  \centering
  \includegraphics[width=\linewidth]{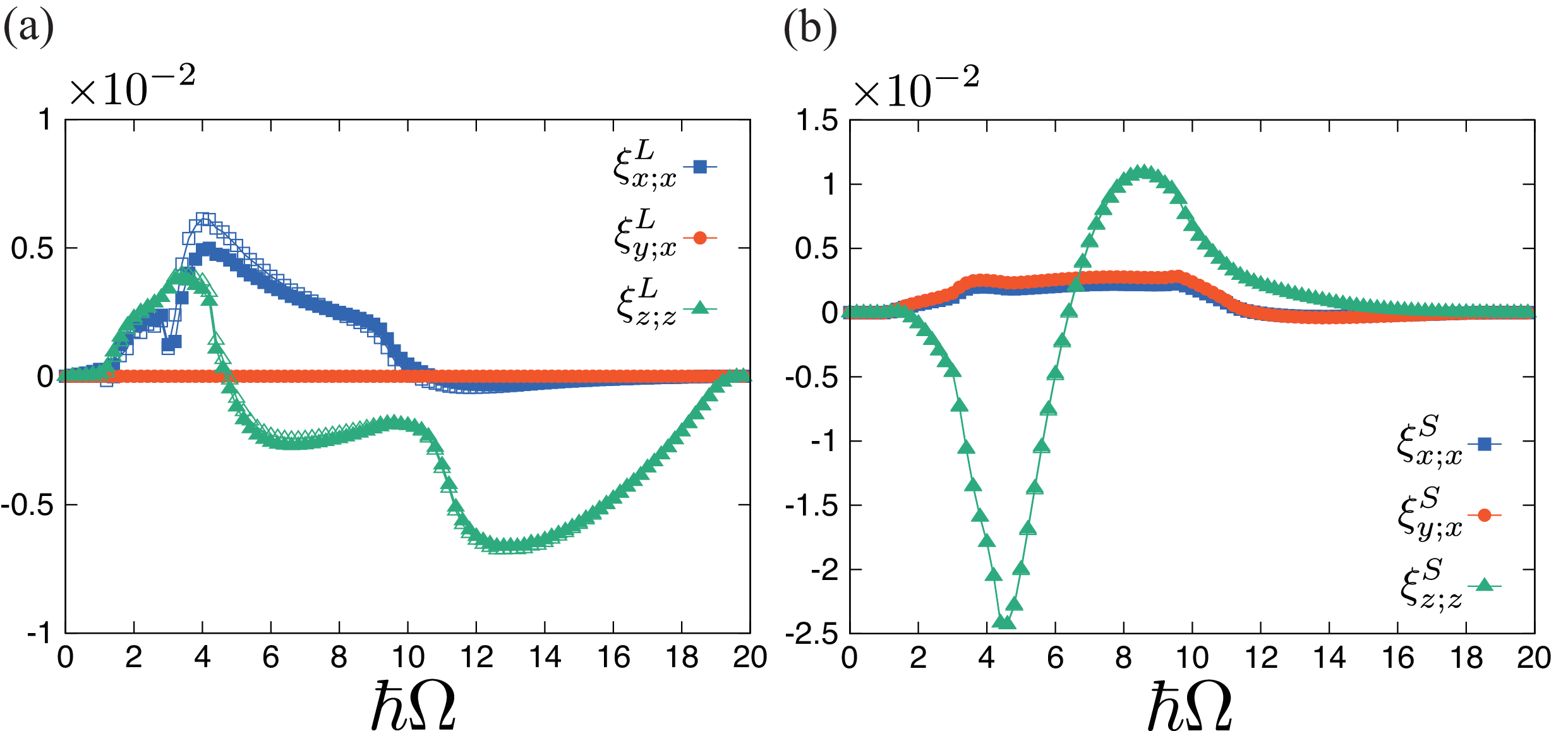}
  \caption{Frequency dependence of the CPL-induced NLEE tensor including the reactive contribution in Eq.~\eqref{eq:gyro_spin}.
  For comparison, the results from Fig.~\ref{fig:nlee_frequency} are also plotted (open).
  We set the same model parameters as those in Fig.~\ref{fig:nlee_frequency}.}
  \label{fig:cp_reactive}
\end{figure*}

We also evaluate the reactive part in Eq.~\eqref{eq:gyro_spin}, which is neglected in the main text.
Figures~\ref{fig:cp_reactive}(a) and (b) show the frequency dependence of total contributions of $\xi_{\mu;\rho}$ using the same model parameters as in Fig.~\ref{fig:nlee_frequency}, that is, $\gamma^{-1}=10^{-2}$.
We also plotted the results from Figs.~\ref{fig:nlee_frequency}(c) and \ref{fig:nlee_frequency}(d) for comparison.
Although the peak strengths are modified, the spectral shapes are similar to results in the main text.
Therefore, ignoring the reactive part is justified in our model calculation in the optical regime.
On the other hand, the reactive part is independent of the scattering rate and becomes nonzero in the low-photon-energy region or even under no-absorption conditions.
Thus, if the scattering is sufficiently large, the reactive part is dominant.
We also note that the reactive part in Eq.~\eqref{eq:gyro_spin} is essentially the three band process.
Since it does not need the orbital polarization of bands, the induced orbital magnetization becomes nonzero even in the absence of the atomic SOC ($\lambda=0$) in our model.

\bibliography{manuscript.bbl}

\end{document}